\documentclass[%
 reprint,
superscriptaddress,
 amsmath,amssymb,
 aps,
prc,
]{revtex4-2}

\usepackage{graphicx}
\usepackage{dcolumn}
\usepackage{bm}
\usepackage[pdfencoding=auto, psdextra]{hyperref}
\usepackage{colortbl}
\usepackage{verbatim}
\usepackage{nicefrac}
\usepackage{comment} 
\usepackage{subcaption}
\usepackage{multirow}
\usepackage{tensor}
\usepackage{footnote}
\usepackage{nicematrix,xcolor}
\usepackage{siunitx}

\newcommand{\rn}[2]{$^{#1}$#2}
\newcommand{\dEc}[1]{$\Delta E_{C}^{(#1)}$}
\newcommand{\dEcb}{$\Delta E_{C}$~}
\newcommand{\ga}{$g_A$~}
\newcommand{\gafree}{$g_A^{\text{free}}$~}
\newcommand{\gainf}{$g_A^{\text{inf}}$~}
\newcommand{\gaeff}{$g_A^{\text{eff}}$~}

\definecolor{burgundy}{rgb}{0.5, 0.0, 0.13}
\definecolor{grannysmithapple}{rgb}{0.66, 0.89, 0.63}

\begin{document}

\preprint{APS/123-QED}

\title{High precision measurement of the \texorpdfstring{\rn{99}{Tc}}{99Tc} \texorpdfstring{$\beta$}{beta} spectrum}

\author{M.~Paulsen}
\email{michael.paulsen@ptb.de}
\affiliation{%
Physikalisch-Technische Bundesanstalt Berlin, Abbestraße 2-12, 10587 Berlin, Germany
}%
\affiliation{%
Kirchhoff-Institute for Physics, Im Neuenheimer Feld 227, 69120 Heidelberg, Germany
}%

\author{P.~C.-O. Ranitzsch}
\altaffiliation{present address:  Deutsches Zentrum für Luft-und Raumfahrt e.V. Hamburg, Hamburg, Germany}
\affiliation{%
Physikalisch-Technische Bundesanstalt Braunschweig, Bundesallee 100, 38116 Braunschweig, Germany
}%

\author{M.~Loidl}
\affiliation{%
Universit{\'e} Paris-Saclay, CEA, List, Laboratoire National Henri Becquerel (LNE-LNHB), F-91120 Palaiseau, France
}%

\author{M.~Rodrigues}
\affiliation{%
Universit{\'e} Paris-Saclay, CEA, List, Laboratoire National Henri Becquerel (LNE-LNHB), F-91120 Palaiseau, France
}%

\author{K.~Kossert}
\affiliation{%
Physikalisch-Technische Bundesanstalt Braunschweig, Bundesallee 100, 38116 Braunschweig, Germany
}%

\author{X.~Mougeot}
\email{xavier.mougeot@cea.fr}
\affiliation{%
Universit{\'e} Paris-Saclay, CEA, List, Laboratoire National Henri Becquerel (LNE-LNHB), F-91120 Palaiseau, France
}%

\author{A.~Singh}
\altaffiliation{present address: Physikalisch-Technische Bundesanstalt Braunschweig, Bundesallee 100, 38116 Braunschweig, Germany }
\affiliation{%
Universit{\'e} Paris-Saclay, CEA, List, Laboratoire National Henri Becquerel (LNE-LNHB), F-91120 Palaiseau, France
}%

\author{S. Leblond}
\affiliation{%
Universit{\'e} Paris-Saclay, CEA, List, Laboratoire National Henri Becquerel (LNE-LNHB), F-91120 Palaiseau, France
}%

\author{J.~Beyer}
\affiliation{%
Physikalisch-Technische Bundesanstalt Berlin, Abbestraße 2-12, 10587 Berlin, Germany
}%

\author{L.~Bockhorn}
\altaffiliation{present address: Institut für Festkörperphysik, Leibniz Universität Hannover, Appelstraße 2, 30167 Hannover, Germany}
\affiliation{%
Physikalisch-Technische Bundesanstalt Braunschweig, Bundesallee 100, 38116 Braunschweig, Germany
}%

\author{C.~Enss}
\affiliation{%
Kirchhoff-Institute for Physics, Im Neuenheimer Feld 227, 69120 Heidelberg, Germany
}%
\author{M.~Wegner}
\affiliation{%
Institute for Data Processing and Electronics (IPE), Karlsruhe Institute of Technology (KIT), Hermann-von-Helmholtz-Platz 1, 76344 Eggenstein-Leopoldshafen, Germany
}%
\affiliation{%
Institute of Micro- and Nanoelectronic Systems (IMS), Karlsruhe Institute of Technology (KIT), Hertzstraße 16, 76187 Karlsruhe, Germany
}%

\author{S.~Kempf}
\affiliation{%
 Institute of Micro- and Nanoelectronic Systems (IMS), Karlsruhe Institute of Technology (KIT), Hertzstraße 16, 76187 Karlsruhe, Germany
 }%
\affiliation{%
Institute for Data Processing and Electronics (IPE), Karlsruhe Institute of Technology (KIT), Hermann-von-Helmholtz-Platz 1, 76344 Eggenstein-Leopoldshafen, Germany
}%
\author{O.~N\"{a}hle}
\affiliation{%
Physikalisch-Technische Bundesanstalt Braunschweig, Bundesallee 100, 38116 Braunschweig, Germany
}%

\date{\today}

\begin{abstract}
\noindent Highly precise measurements of the $^{99}$Tc beta spectrum were performed in two laboratories using metallic magnetic calorimeters. Independent sample preparations, evaluation methods and analyses yield consistent results and the spectrum could be measured down to less than \SI{1}{\kilo\electronvolt}. Consistent beta spectra were also obtained via cross-evaluations of the experimental data sets. An additional independent measurement with silicon detectors in a $4\pi$ configuration confirms the spectrum shape above \SI{25}{\kilo\electronvolt}. Detailed theoretical calculations were performed including nuclear structure and atomic effects. The spectrum shape was found to be sensitive to the effective value of the axial-vector coupling constant. Combining measurements and predictions, we extracted $Q_{\beta} =$\SI{295.82 (16)}{\kilo\electronvolt} and $g_A^{\text{eff}} = 1.530 (83)$. Furthermore, we derived the mean energy of the beta spectrum $\overline{E}_{\beta}$=\SI{98.45(20)}{\kilo\electronvolt}, $\log f = -0.47660 (22)$ and $\log ft = 12.3478 (23)$.
\end{abstract}

\maketitle

\section{Introduction}
Beta spectrometry, the precise shape of beta spectra and their theoretical description have received increased interest recently from different research fields, e.g. radionuclide metrology \cite{Kos15, Kossert2022}, neutrino physics \cite{Mou15, brdar2022empirical} and nuclear theory \cite{Mou14,Kostensalo2017}. In the context of radionuclide metrology, the European metrology research project MetroBeta \cite{Loidl18} (2016-2019) addressed the precise measurement and theoretical calculation of several beta spectra.

The ground state of \rn{99}{Tc} decays mainly via pure beta emission ($\beta^{-}$, 99.99855(30)\%) to the \rn{99}{Ru} ground state \cite{DDEP_v6}. The spectrum shape of this $2^{\mathrm{nd}}$ forbidden non-unique transition has been measured several times using magnetic \cite{Feldman52,Taimuty51}, scintillation \cite{Sny66} and semiconductor \cite{Reich74} spectrometers \cite{Behrens76}. While these setups corresponded to the state-of-the-art when they were applied in the 1950s-70s, the measurements suffer from rather high energy thresholds ($>$ \SI{50}{\kilo\electronvolt}) and it is expected that one can achieve significantly higher energy resolution with present-day methods. In addition, the beta spectrum shape of \rn{99}{Tc} has recently been predicted in \cite{Kostensalo2017} to be very sensitive to the effective value of the weak interaction axial-vector coupling constant $g_A$, making its high-precision measurement very interesting. It is also expected that the currently assumed $Q$-value of \SI{297.5(9)}{\kilo\electronvolt} \cite{Wang2021}, which has a relative standard uncertainty of about 0.3\%, can be determined more accurately with modern methods. Recently, the influence of nuclear data on decay heat from spent nuclear fuel over a period of 1 to 100k years was assessed \cite{Doran2022}. A list of the most significant contributing radionuclides was provided. \rn{99}{Tc} was placed at the very top, with an average beta energy that ranges from 84 to \SI{95}{\kilo\electronvolt} and a stated uncertainty of less than 1\%, depending on the data library.\newline
\indent In the framework of the MetroBeta project, a beta spectrum of \rn{99}{Tc} was measured with Metallic Magnetic Calorimeters (MMCs), which was first presented in \cite{Loidl19,Loidl20}. It featured two orders of magnitude lower energy thresholds (\SI{0.65}{\kilo\electronvolt}) and a greatly improved energy resolution (\SI{0.1}{\kilo\electronvolt} at \SI{383}{\kilo\electronvolt}) compared to previous measurements. The spectrum was obtained at the Laboratoire National Henri Becquerel (LNHB) and shows excellent agreement with a corresponding novel measurement at the Physikalisch-Technische Bundesanstalt (PTB) using a similar MMC setup, which confirms the spectrum shape over the entire energy range. At energies above \SI{25}{\kilo\electronvolt} the spectrum shapes are further confirmed with a state-of-the-art Passivated Implanted Planar Silicon (PIPS) detector measurement, also performed at LNHB.\newline
\indent In this work, we report on these three independent measurements of the \rn{99}{Tc} beta spectrum and on the data analyses performed to correct for small, but relevant distortions due to the detection systems. We next present a spectrum analysis that combines these accurate measurements with detailed theoretical predictions in order to extract the \rn{99}{Tc} $Q$-value and the effective $g_A$ coupling constant. The average energy of the beta spectrum and the $\log ft$ value have also been derived.



\section{Experimental study}
\label{sec:experimental_study}
\subsection{MMC measurements}
\label{subsec:MMC}
MMCs \cite{Fleischmann05, Fleischmann09, Kempf18} are cryogenic microcalorimeters that consist of a - mostly metallic - particle absorber in strong thermal contact with a metallic paramagnet (here: Ag:Er$_{300\,\mathrm{ppm}}$) acting as a temperature sensor. The paramagnet is placed in a weak magnetic field ($\sim$\SI{10}{\milli\tesla}) and operated at temperatures $<$ \SI{100}{\milli\kelvin}. When an energy $E$ is deposited into the absorber, it leads to a temperature increase $\Delta T$. As the magnetic susceptibility of the paramagnet has a strong temperature dependence,  the temperature increase causes a change in its magnetization
\begin{equation}
\Delta M = \frac{\partial M}{\partial T}\cdot\Delta T = \frac{\partial M}{\partial T}\cdot\frac{E}{C_{\text{tot}}},
\end{equation}
where $C_{\text{tot}}$ denotes the total heat capacity of the absorber and the paramagnet. A superconducting coil coupled to the paramagnet picks up the change in magnetization as a corresponding magnetic flux change $\Delta \Phi$, which is measured with a Superconducting QUantum Interference Device (SQUID) \cite{Drung07}.

Setups using MMCs with the radionuclide source embedded in a 4$\pi$ solid angle absorber geometry have proven to be among the best beta spectrometers in terms of energy resolution and energy threshold, in particular for low-energy beta transitions \cite{rotzinger2008beta,Loidl14,Loidl19,Loidl20}. Both measurements presented here follow that approach, but the technical realization differs in many details. These are summarized in Table~\ref{tab:MMC_LNHB_vs_PTB} and described in the following. 

\begin{table}[htbp]
\caption{Setup properties of the MMC experiments at the LNHB and PTB laboratories. The values for the input inductance and heat capacity are nominal per fabrication and calculated values, respectively.}
  \begin{center}
\resizebox{\columnwidth}{!}{
    \begin{NiceTabular}{|l|l|l|}[vlines]
   \Hline
 \textbf{Laboratory} & \Block[fill=grannysmithapple!20]{1-1}{\textbf{LNHB} \cite{Loidl18,Loidl19,Loidl20}} & \Block[fill=burgundy!30]{1-1}{\textbf{PTB} \cite{Paulsen19,PaulsenPhD2022}}\\
\Hline
     & $^{3/4}$He dilution insert & $^{3/4}$He dilution refrigerator\\
    \textbf{Cryostat} & in $^{4}$He~(l) bath & with two stage pulse tube\\
    &  for pre-cooling  & for pre-cooling\\
    \Hline
     \textbf{MMC chip} & MetroBeta V1-M  \cite{Loidl18} & MetroBeta V2-M \cite{PaulsenPhD2022}\\
    \Hline
     \textbf{SQUID chip} & Supracon VC1A & PTB X1 \\
    & input inductance \SI{4.5}{\nano\henry} & input inductance  \SI{2}{\nano\henry}\\
    \Hline
      & Au, heat capacity & Au, heat capacity \\
     \textbf{Absorber} & \SI{350}{\pico\joule/\kelvin} at \SI{20}{\milli\kelvin} & \SI{112}{\pico\joule/\kelvin} at \SI{20}{\milli\kelvin}\\
     & matched to MMC chip & matched to MMC chip\\
    \Hline
     \textbf{Sample} & electrodeposited & drop deposited \\
    \Hline
    \textbf{Calibration source} & $^{133}$Ba & $^{57}$Co \\
    \Hline
     \textbf{Analysis code} & Optimal filtering & Optimal filtering\\
     & in MATLAB\textsuperscript{\textregistered}  & in Python \\
    \Hline
    \end{NiceTabular}
}
\end{center}
  \label{tab:MMC_LNHB_vs_PTB}
\end{table}%

\subsubsection{Setup and analysis (LNHB)}
The starting point of the detector fabrication at LNHB was the source preparation. Following a protocol yielding metallic technetium \cite{mausolf2011characterization}, \rn{99}{Tc} was electrodeposited onto a \SI{10}{\micro\meter} thick gold foil. The foil was then rinsed with water in order to remove salt having crystallized from the \rn{99}{Tc} solution on the foil. Some visible salt deposit remained even after rinsing, but in an autoradiographic image of the source several areas without any salt deposit but with presence of \rn{99}{Tc} activity were found. The electrodeposition yield was low and an area of the source foil larger than the typical size of MMC absorbers had to be used to have sufficient activity in the MMC absorber. The selected piece of source foil ($\sim$\SI{0.9}{\milli\meter}$\times$\SI{2.5}{\milli\meter}) with a transparently thin metallic \rn{99}{Tc} deposit was folded three times to reduce it to a small enough size ($\sim$\SI{0.44}{\milli\meter}$\times$\SI{0.64}{\milli\meter}$\times$\SI{54}{\micro\meter}) such that it could be enclosed into the MMC absorber. The folded foil with a \rn{99}{Tc} activity of $\sim$~\SI{5}{\becquerel} was sandwiched between two gold foils (\SI{0.9}{\milli\meter}$\times$\SI{0.9}{\milli\meter}$\times$\SI{74}{\micro\meter} each) and this stack was diffusion welded. The final absorber had a heat capacity of $\sim$ \SI{350}{\pico\joule/\kelvin} at \SI{20}{\milli\kelvin} and was glued with Stycast 1266 epoxy  to one of the pixels of a MetroBeta V1 M-sized MMC chip \cite{Loidl18}.

Due to the continuous nature of the beta spectra, without any distinct features, energy calibration is essential, in particular if the end point energy is to be determined from an experimental spectrum. To precisely determine the spectrum shape, checking and correcting for any nonlinearities in energy is also important. A common way to perform energy calibration is to use X-ray and/or gamma ray photons of well-known energies from an external radionuclide source collimated onto the detector. To cover the full energy range of the $\mathrm{^{99}Tc}$ beta spectrum, at LNHB a $\mathrm{^{133}Ba}$ source was chosen. It emits X-ray and gamma lines between \SI{30.63}{\kilo\electronvolt} and \SI{383.85}{\kilo\electronvolt}. It was placed at a distance of \SI{31}{\milli\meter} from the absorber surface. A lead collimator was composed of a \SI{8}{\milli\meter} thick top part with a \SI{1}{\milli\meter} bore and a \SI{2}{\milli\meter} thick bottom part with a \SI{200}{\micro\meter} bore placed at \SI{2.5}{\milli\meter} from the absorber surface. The MMC signal was read out by a Supracon VC1A SQUID linked to a Magnicon XXF-1 electronics.

The whole setup was shielded against stray magnetic fields by means of a lead cylinder and operated in a liquid helium pre-cooled $^{3/4}$He dilution refrigerator (Cryoconcept) at $T=$ \SI{12}{\milli\kelvin}. Data was acquired as a continuous stream over 13.7\,days at 100\,kS/s; anti-aliasing filtering was set to \SI{30}{\kilo\hertz} on a Stanford Research Systems SRS 560 amplifier.

\begin{figure}[ht]
    \begin{center}
        \includegraphics[width=\columnwidth]{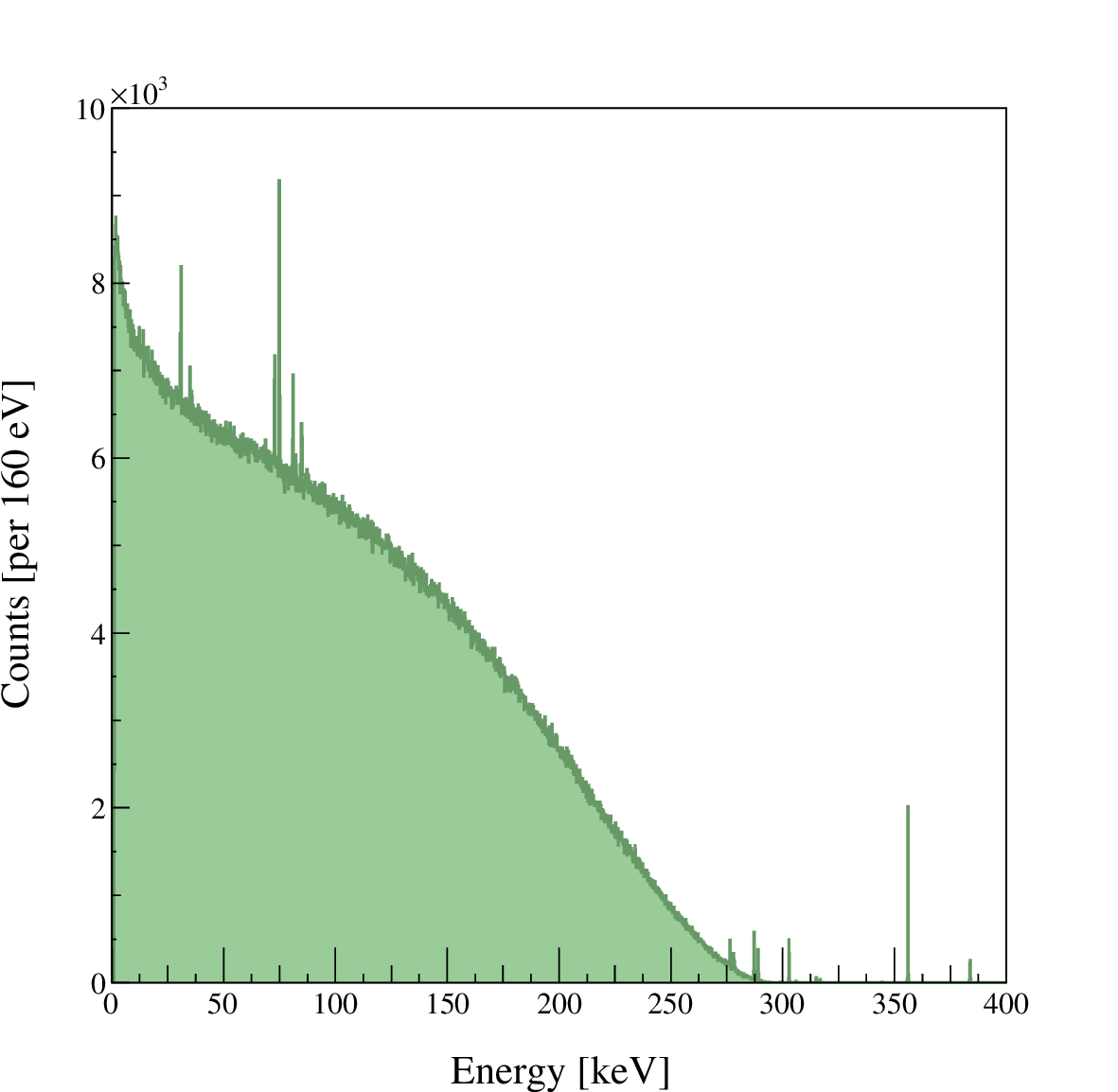}
        \caption{Measured and calibrated LNHB MMC spectrum of \rn{99}{Tc} with \rn{133}{Ba} calibration peaks.}
        \label{Fig:LNHB_measured_calibrated}
    \end{center}
\end{figure}

Pulses were triggered in the data stream offline, using a narrow band pass filter. To minimize pile-up, an extendable dead-time was applied. Once the pulse positions were determined, the pulse heights were estimated from raw data using an optimal filter in a MATLAB\textsuperscript{\textregistered} environment. Slow variations of the pulse heights for a given energy due to temperature drifts of the cryostat were removed by fitting the pulse height-vs-time distribution for one densely populated line with a spline function and applying this fit function to all pulse heights. Spurious pulses were discriminated based on the pulse shapes in a chi-square-vs-pulse-height plot. The final spectrum contains 7264451\,events and is presented in Fig. \ref{Fig:LNHB_measured_calibrated}.

The energy resolution is constant over the entire energy range, $\Delta E\approx$  \SI{100}{\electronvolt} up to \SI{384}{\kilo\electronvolt}. The linearity in energy was checked using the lines listed in Table  \ref{tab:linearity_check_calibration}. First the energy scale was set using uniquely the \SI{356.01}{\kilo\electronvolt} line, an intense gamma line of the $\mathrm{^{133}Ba}$ calibration source lying beyond the endpoint of the $\mathrm{^{99}Tc}$ beta spectrum. Then the resulting energies of the other lines from the MMC measurement were compared with the recommended energies \cite{DDEP_v1}, where for the escape lines the X-ray transition energies from \cite{deslattes2003x} were used. The differences in energy between tabulated and measured energies are all smaller or equal to the energy resolution; for the \SI{30.97}{\kilo\electronvolt} line, the most intense X-ray line, it is zero. There is no clear trend in the difference between tabulated and measured energies as a function of energy, but it could be described with a second order polynomial. This polynomial was then used to correct the energy scale for the non-linearity.

\begin{table*}[ht]
\centering
  \caption{Photon and escape lines used to check and correct for energy non-linearity.}
  \resizebox{\textwidth}{!}{
  \bgroup
\def\arraystretch{1.1}%
\begin{NiceTabular}{|c|c|c|c|c|c|c|}
\Hline
\Block[fill=grannysmithapple!20]{1-7}{\textbf{LNHB measurement}}\\
\Hline
            \textbf{ } & \textbf{Tabulated} & \textbf{Measured} & \textbf{$\Delta E$ [Set 3]} & \multicolumn{2}{l|}{\textbf{ }} & \textbf{Calibration line} \\
    \textbf{Origin} & \textbf{energy in keV}  & \textbf{energy in keV} & \textbf{in keV} & \multicolumn{2}{c|}{\textbf{Radiation type }} & \textbf{for Set}\\
\Hline
    \textbf{$^{133}$Ba calibration} & 30.63    & 30.61 & -0.02    & \multicolumn{2}{c|}{Cs X K$_{{\alpha}2}$}   & 3,4\\
    \textbf{source} & 30.97    & 30.97 & 0    & \multicolumn{2}{c|}{Cs X K$_{{\alpha}1}$}   & 3,4\\
    \textbf{ } & 35.05    & 34.98 & -0.07    & \multicolumn{2}{c|}{Cs X K$_{\beta}$}   & -\\
    \textbf{ } & 81.00    & 81.06 & 0.06    & \multicolumn{2}{c|}{$\gamma$} & -\\
    \textbf{ } & 276.40    & 276.46 & 0.06    & \multicolumn{2}{c|}{$\gamma$} & -\\
    \textbf{ } & 302.85    & 302.88 & 0.00    & \multicolumn{2}{c|}{$\gamma$} & 3,4\\
    \textbf{ } & 356.01    & 356.01 & 0       & \multicolumn{2}{c|}{$\gamma$} & 3,4\\
    \textbf{ } & 383.85    & 383.82 & -0.03   & \multicolumn{2}{c|}{$\gamma$} & 3,4\\ \Hline
    \textbf{Au/Pb fluorescence} & 66.99   & 66.98 & -0.01 & \multicolumn{2}{c|}{Au X K$_{{\alpha}2}$} & -\\
    \textbf{from}                            & 68.80   & 68.78 & -0.02 & \multicolumn{2}{c|}{Au X K$_{{\alpha}1}$} & -\\
    \textbf{setup/collimator}                            & 72.80   & 72.82 & 0.02 & \multicolumn{2}{c|}{Pb X K$_{{\alpha}2}$} & 3,4\\
     & 74.97   & 74.98 & 0.01 & \multicolumn{2}{c|}{Pb X K$_{{\alpha}1}$} & 3,4\\
     & 84.94   & 84.98 & 0.04 & \multicolumn{2}{c|}{Pb X K$_{{\beta}1}$} & -\\
    \Hline
    \textbf{ } &    &  &  & \multicolumn{1}{p{8em}|}{\textbf{Gamma line in keV}} & \multicolumn{1}{p{6em}|}{\textbf{Escaping photon in keV}} & \\
     \Hline
    \textbf{Escape lines} & 278.03    & 278.07 & 0.04 & 356.01 [$^{133}${Ba} $\gamma$] & 77.98 [Au K$_{{\beta}1}$] & -\\
    \textbf{ } & 278.43    & 278.54 & 0.11   & 356.01 [$^{133}${Ba} $\gamma$] & 77.58 [Au K$_{{\beta}3}$] & -\\
    \textbf{ } & 287.21    & 287.28 & 0.07   & 356.01 [$^{133}${Ba} $\gamma$] & 68.80 [Au K$_{{\alpha}1}$] & 3,4\\
    \textbf{ } & 289.02    & 289.08 & 0.06   & 356.01 [$^{133}${Ba} $\gamma$] & 66.99 [Au K$_{{\alpha}2}$] & 3,4\\
    \textbf{ } & 305.86    & 305.92 & 0.06   & 383.85 [$^{133}${Ba} $\gamma$] & 77.98 [Au K$_{{\beta}1}$] & -\\
    \textbf{ } & 315.05    & 315.08 & 0.03   & 383.85 [$^{133}${Ba} $\gamma$] & 68.80 [Au K$_{{\alpha}1}$] & -\\
    \textbf{ } & 316.85    & 316.91 & 0.05   & 383.85 [$^{133}${Ba} $\gamma$] & 66.99 [Au K$_{{\alpha}2}$] & -\\
\Hline
\Block[fill=burgundy!30]{1-7}{\textbf{PTB measurement}}\\
\Hline
            \textbf{ } & \textbf{Tabulated} & \textbf{Measured} & \textbf{$\Delta E$ [Set 2]} & \multicolumn{2}{l|}{\textbf{ }} & \textbf{Calibration line}\\
    \textbf{Origin} & \textbf{energy in keV}  & \textbf{energy in keV} & \textbf{in keV} & \multicolumn{2}{c|}{\textbf{Radiation type }} & \textbf{for Set}\\
\Hline
    \textbf{$^{57}${Co} calibration} & 14.42    & 14.44 & -0.02    & \multicolumn{2}{c|}{$\gamma$}    & -\\
    \textbf{source} & 122.06    & 122.06 & 0.00    & \multicolumn{2}{c|}{$\gamma$}    & 1,2\\
    \textbf{ } & 136.47    & 136.38 & 0.09    & \multicolumn{2}{c|}{$\gamma$}    & 1,2\\
\Hline
 \textbf{Au/Pb fluorescence} & 66.99   & 67.07 & -0.08 & \multicolumn{2}{c|}{Au X K$_{{\alpha}2}$} & -\\
    \textbf{from}            & 68.80   & 68.95 & -0.15 & \multicolumn{2}{c|}{Au X K$_{{\alpha}1}$} & -\\
\textbf{setup/collimator} & 72.81   & 72.98 & -0.17 & \multicolumn{2}{c|}{Pb X K$_{{\alpha}2}$} & 1,2\\
     & 74.97   & 75.14 & -0.17 & \multicolumn{2}{c|}{Pb X K$_{{\alpha}1}$} & 1,2\\
     & 84.94   & 85.09 & -0.15 & \multicolumn{2}{c|}{Pb X K$_{{\beta}1}$} & -\\
    \Hline
    \textbf{ } &    &  &  & \multicolumn{1}{p{8em}|}{\textbf{Gamma / fluorescent line in keV}} & \multicolumn{1}{p{6em}|}{\textbf{Escaping photon in keV}} & \\
     \Hline
\textbf{Escape lines} & 6.47    & 6.44 & 0.03   & 84.45 [Pb K$_{{\beta}3}$] & 77.98 [Au K$_{{\beta}1}$] & -\\
    \textbf{ } & 6.96    & 6.97 & -0.01   & 84.94 [Pb K$_{{\beta}1}$] & 77.98 [Au K$_{{\beta}1}$] & -\\
    \textbf{ } & 16.13    & 16.22 & -0.08   & 84.94 [Pb K$_{{\beta}1}$] & 68.80 [Au K$_{{\alpha}1}$] & 1,2\\
    \textbf{ } & 53.26    & 53.42 & -0.17   & 122.06 [$^{57}${Co} $\gamma$] & 68.80 [Au K$_{{\alpha}1}$] & 1,2\\
    \textbf{ } & 55.07    & 55.24 & -0.17   & 122.06 [$^{57}${Co} $\gamma$] & 66.99 [Au K$_{{\alpha}2}$] & 1,2\\
\Hline
\end{NiceTabular}
\egroup
}
\label{tab:linearity_check_calibration}
\end{table*}

\subsubsection{Setup and analysis (PTB)}
The measurement setup at PTB is functionally the same as at LNHB, but differed in several details. The difference with possibly the largest influence on the measured spectrum shape is the method of source preparation. This was done by micro-dispensing an  aqueous solution of ammonium pertechnetate ($\mathrm{NH_{4}{}^{99}TcO_{4}}$) in \SI{0.1}{\mol\per\liter} ammonia ($\mathrm{NH_{3}}$) with an activity of $A(\mathrm{^{99}Tc})\approx$~\SI{5}{\becquerel} onto a solid gold absorber substrate with a thickness of \SI{90}{\micro\meter}. The source is enclosed by diffusion welding a second \SI{90}{\micro\meter} gold layer to the first one, with more details of the source preparation process being described in~\cite{Bockhorn2020}. The absorber is attached to one pixel of the MMC detector with Stycast 1266 epoxy and the second pixel is equipped with a second absorber prepared in the same way but without any radioactive source material. The choice of photon calibration source also has an impact on the spectrum shape and $^{57}$Co was used for the measurement at PTB. A \SI{1.5}{\milli\meter} thick lead collimator with two \SI{250}{\micro\meter} apertures blocks the calibration photons outside of the two detector pixels. The apertures are blocked between detector and collimator with an approximately \SI{1}{\milli\meter} thick aluminium sheet to reduce secondary radiation from the lead without significantly impacting the high-energy photon flux from the $^{57}$Co source.\newline
\indent The detector was operated in a pulse-tube pre-cooled $^{3/4}$He dilution refrigerator (Bluefors LD250) temperature stabilized to $T=$~\SI{14.5}{\milli\kelvin} on a detector module described in~\cite{Loidl18}. The used MMC (size M) is an update to the one described in the same publication, with some layout changes improving on experimental shortcomings observed with the first design (e.g. the two on-chip heat baths were linked in the new design to improve thermalization), without changing the core properties of the detector. The setup is completed with a PTB SQUID (model C6X114W) for MMC read-out, a Magnicon XXF-1 SQUID electronics, a Stanford Research Systems SR560 low-noise voltage pre-amplifier and band pass filter, and a 16-Bit waveform digitizer, that was set to save the full data stream at 200\,$\mathrm{kS\cdot s^{-1}}$ to hard disc for the measurement's duration of about 20\,days.
\begin{figure}[ht]
    \begin{center}
        \includegraphics[width=\columnwidth]{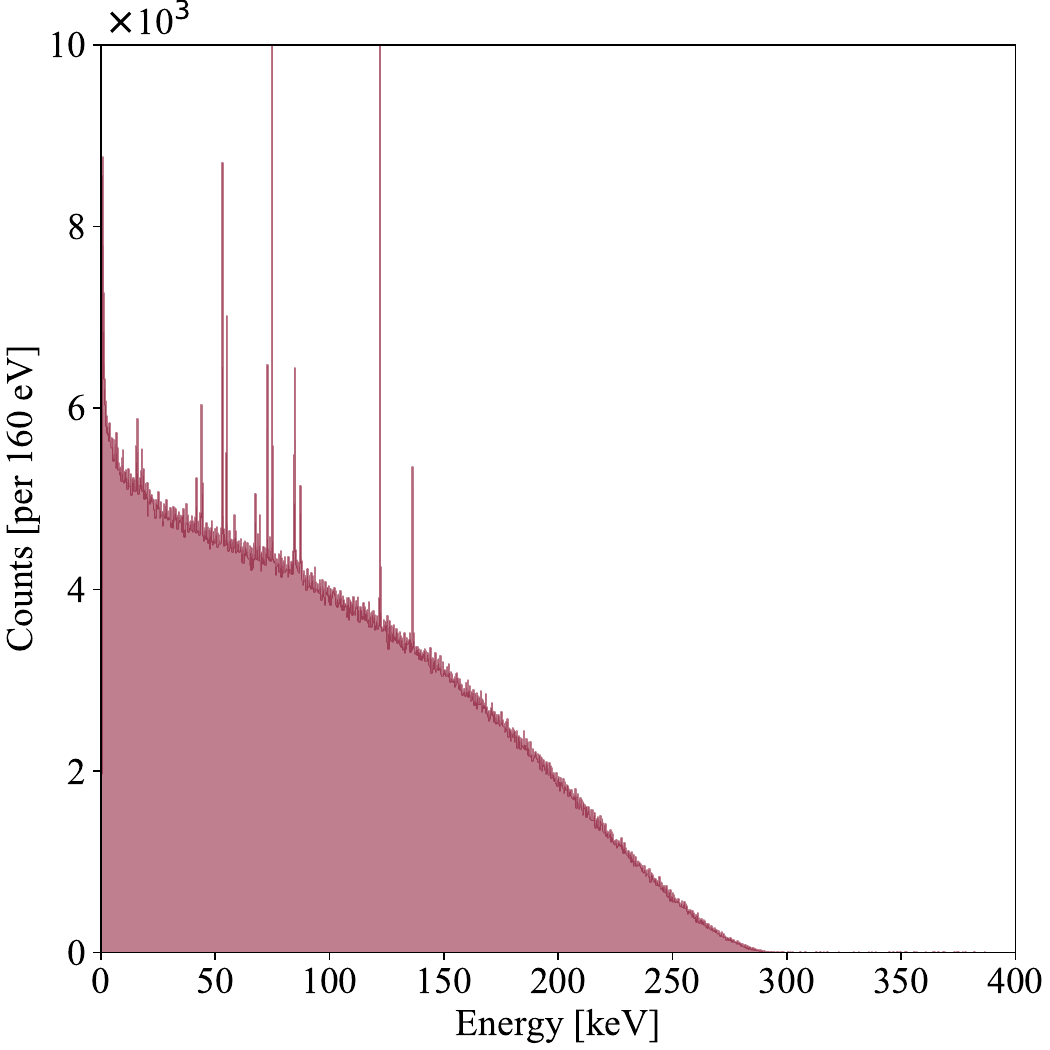}
        \caption{Measured and calibrated PTB MMC spectrum of \rn{99}{Tc} with \rn{57}{Co} calibration peaks.}
        \label{Fig:PTB_measured_calibrated}
    \end{center}
\end{figure}
The data processing and analysis are performed with a custom software code written in Python, that performs the next steps which comprise event triggering, amplitude/energy determination, event classification through pulse shape discrimination and gain drift corrections. The analysis yields 5326682\,counts in the spectrum after cuts. The resulting amplitude distribution is calibrated with several $\mathrm{^{57}Co}$ $\gamma$-lines, X-ray fluorescence and X-ray escape lines using a second order polynomial, returning the spectrum shown in Fig. \ref{Fig:PTB_measured_calibrated}. The calibration error was estimated with the uncertainties from the literature values and the statistical uncertainties of the measured peaks using Orthogonal Distance Regression \cite{PaulsenPhD2022,Bog90,Bat16,Odr21}. The statistical uncertainty was defined as $\sigma_{\text{Gauss}}/\sqrt{N_{\text{peak}}}$, where $\sigma_{\text{Gauss}}$ denotes the standard deviation of a Gaussian function that was fitted to the peak and $N_{\text{peak}}$ is the number of counts in the peak. The same procedure was done with the pixel without radioactive source material yielding a spectrum containing the $\mathrm{^{57}Co}$ calibration spectrum and any additional background. There a line spectrum with 132235\,counts was extracted. 

\subsubsection{Cross analysis}
In both institutes the ADC signals were directly streamed to hard disc and data were saved as binary files containing 16-bit integers. Therefore, the raw data of both measurements are compatible with each other. This prompted an exchange of the data and a mutual analysis at both institutes to identify any systematic differences introduced by the data processing code or analysis approach. The data processing and analysis were conducted as usual for local data sets and allowed for the comparison between the four spectra in total resulting from two measurements with two analyses each.

For both data sets, the two separate analyses yield surprisingly different numbers of counts, as can be seen in Table~\ref{tab:Cross_Analysis_Counts_Energy_Resolutions}. The PTB measurement involuntarily stopped during the campaign and was restarted. This caused inconsistencies in the data and for the analysis at PTB, successful corrections were implemented. In contrast, the LNHB analysis only used the largest continuous data set, that contained about three-quarters of the total data. The remaining difference, also for the LNHB measurement, is most likely caused by differences in the implementation and the settings for the software trigger and possibly by cuts for event selection. The trigger threshold should only influence the energy threshold, but other settings, such as trigger holdoff or extending/non-extending dead-time etc. can have an impact over the whole energy range. Specifically for the LNHB-LNHB [set 3] analysis, a very short event time window was chosen, that allows to reduce the impact of pile-up, with the trade-off of a slightly diminished energy resolution and higher energy threshold. \newline
\indent It was agreed upon to use the same calibration lines for the energy calibration in the spectra and the corresponding literature values for the $\gamma$ \cite{DDEP_v8,DDEP_v9} and X-ray \cite{deslattes2003x} energies. These lines were very easy to identify and to fit in the spectrum. The tabulated energies and differences to the measurement data are summarized in Table \ref{tab:linearity_check_calibration}.

\begin{table}[ht!]
\centering
    \caption{Overview of the cross validation with counts, FWHM and energy threshold (=$E_{\text{TH}}$) values. Set 1 only considered a subset of the recorded data.}
    \resizebox{\columnwidth}{!}{
    \bgroup
\def\arraystretch{1.2}
\begin{NiceTabular}{cccccc}[hvlines,corners=NW,colortbl-like]
\bf{Set}  & \bf{Meas.} & \bf{Analysis} & \bf{Counts} & \bf{FWHM} & \bf{$E_{\text{TH}}$}\\
1 & \Block[fill=burgundy!30]{1-1}{PTB} & \Block[fill=grannysmithapple!20]{1-1}{LNHB} & $3.67\cdot 10^{6}$ & \SI{63}{\electronvolt}@\SI{136.47}{\kilo\electronvolt} & \SI{345}{\electronvolt} \\
2 & \Block[fill=burgundy!30]{1-1}{PTB} & \Block[fill=burgundy!30]{1-1}{PTB} & $5.33\cdot 10^{6}$ & \SI{72}{\electronvolt}@\SI{136.47}{\kilo\electronvolt}  & \SI{750}{\electronvolt}\\
3 & \Block[fill=grannysmithapple!20]{1-1}{LNHB} & \Block[fill=grannysmithapple!20]{1-1}{LNHB} & $7.26\cdot 10^{6}$ & \SI{150}{\electronvolt}@\SI{302.8}{\kilo\electronvolt} & \SI{1120}{\electronvolt}\\
4 & \Block[fill=grannysmithapple!20]{1-1}{LNHB} & \Block[fill=burgundy!30]{1-1}{PTB} & $5.66\cdot 10^{6}$ & \SI{108}{\electronvolt}@\SI{302.8}{\kilo\electronvolt}  & \SI{1250}{\electronvolt}\\
\end{NiceTabular}
\egroup
}
\label{tab:Cross_Analysis_Counts_Energy_Resolutions}
\end{table}

\subsubsection{Spectrum corrections}

Before the actual shape of the \rn{99}{Tc} beta spectrum shape can be investigated, two corrections need to be applied to the measured data. Firstly, the contributions from the calibrations sources and additional background need to be removed. Secondly, the spectrum needs to be corrected for energy losses of the beta electrons, mostly caused by X-ray fluorescence of the absorber material and escaping bremsstrahlung.

The simplest way to account for background is to assume a constant background, with its level being determined from the background above the end-point of the beta spectrum. This approach could e.g. be sufficient for the data taken at PTB and is used in its analysis at LNHB, since no major background sources or $\gamma$-lines are expected in the end-point region of the beta spectrum, where background has the strongest impact on the spectrum shape. With this background model, the spectrum shape cannot be evaluated at the position of $\gamma$-lines, which is not a big drawback because of the narrow line widths.

For the measurement at LNHB, this approach needs to be extended, because several calibration lines are close to and even above the beta end-point region. While the background between lines can reasonably be assumed to be constant, the level becomes slightly higher below each line and the background is described as a series of step functions with the step height being proportional to the line intensity. With this approach the background in both analyses can be reasonably well described.

The measurement at PTB also offers a different approach to evaluate the background, which is used in the PTB analysis. The second pixel of the detector is equipped with an absorber without any enclosed radioactive material, which is also illuminated by the calibration source. Therefore, the spectrum of the second pixel should be a very good approximation for the background of the primary pixel. Since the detector performance, e.g. in rise time and energy resolution, is not exactly the same, these need to be adjusted and the amplitude of the measured background spectrum scaled to match the intensity of the primary pixel, before the background is subtracted. Since the primary pixel showed the better energy resolution ($\Delta E_{\mathrm{FWHM}}=$ \SI{72}{\electronvolt} compared to $\Delta E_{\mathrm{FWHM}}=$ \SI{92}{\electronvolt}), the primary spectrum was convolved with a normalized gaussian with $\Delta E_{\mathrm{FWHM}}$ = \SI{28}{\electronvolt} to match the secondary spectrum and the secondary spectrum was scaled by a factor of 0.73 because of its larger intensity.

To account for energy losses in the absorber, mainly via unstopped photons generated by excitation of Au atoms and bremsstrahlung, an unfolding correction was applied to each spectrum before determining the maximum beta energy. The unfolding procedure is based on an algorithm that does not require any a priori knowledge of the true beta spectrum \cite{Paulsen2020,PaulsenPhD2022}. Its basis is the following discrete unfolding problem:

\begin{equation}
    \mathbf{h}_{N}^{\text{meas}}=\mathbf{R}_{N\times N}\cdot\mathbf{h}_{N}^{\text{true}},
    \label{eq:unfolding_problem}
\end{equation}
where $\mathbf{h}_{N}^{\text{meas}}$ is the measured histogram spectrum having $N$ energy bins, $\mathbf{R}_{N\times N}$ denotes the response matrix of the absorber for the $N$ energy intervals and $\mathbf{h}_{N}^{\text{true}}$ is the unknown true histogram spectrum. After a measurement, one only obtains a value for the left-hand side of (\ref{eq:unfolding_problem}) and thus the problem is typically high-dimensional and greatly under-determined. However, an excellent approximation of the response matrix $\mathbf{R}_{N\times N}^{\text{sim}}\approx\mathbf{R}_{N\times N}$ can be acquired via large-scale Monte Carlo (MC) simulations of the corresponding absorber geometries. Since it can be shown that the approximate matrix is invertible with probability 1 [\onlinecite{Paulsen2020}], the algorithm provides an approximate solution to Eq.  (\ref{eq:unfolding_problem}):

\begin{equation}
    \mathbf{h}_{N}^{\text{true}}\approx\mathbf{h}_{N}^{\text{algo}}\overset{!}{=}\left(\mathbf{R}_{N\times N}^{\text{sim}}\right)^{-1}\cdot\mathbf{h}_{N}^{\text{meas}}.
    \label{eq:algo_solution}
\end{equation}

\begin{figure}[htbp]
    \begin{center}
        \includegraphics[width = 0.97\columnwidth]{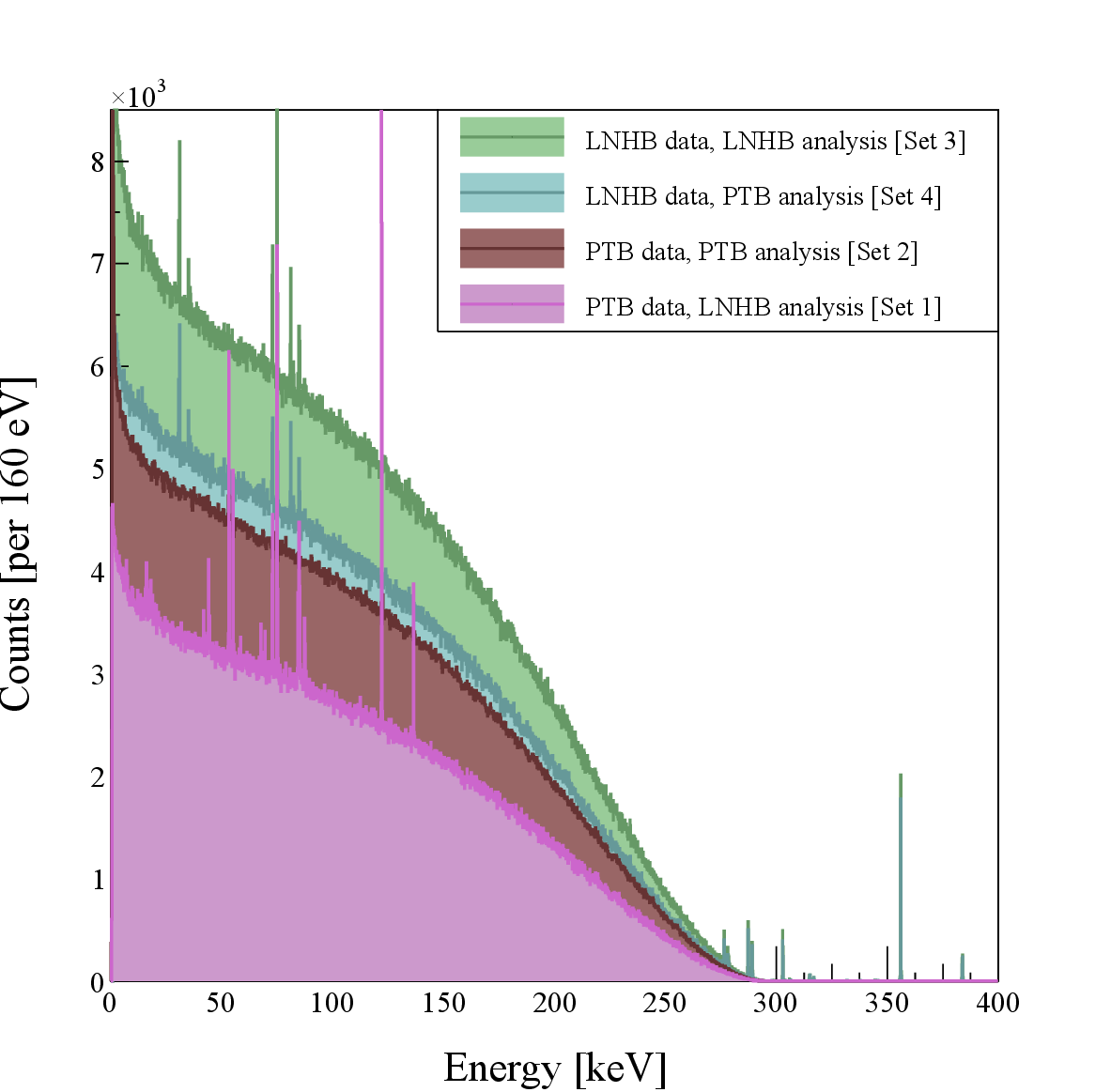}
        \caption{The $^{99}$Tc spectra measured by MMCs, with their separate analyses. Background and energy losses have been corrected as described in the text.}
        \label{Fig:MMCSpectra}
    \end{center}
\end{figure}

The adopted absorber geometries are, notably, just an approximation since the exact distributions of the activity within the rather complex sources are not precisely known. In addition, the imperfect knowledge of the cross sections (e.g. for bremsstrahlung) contribute to the related uncertainty component. The MC simulations were realized with the code egs\_phd within the EGSnrc software \cite{Kaw22} and were performed separately for the PTB and LNHB detector geometries. The resulting four spectra with the background removed and energy loss corrections applied are shown in Fig.~\ref{Fig:MMCSpectra}.

\subsubsection{Inter-comparison}
\begin{figure*}[tbhp]
    \begin{center}
        \includegraphics[width = 0.95\textwidth]{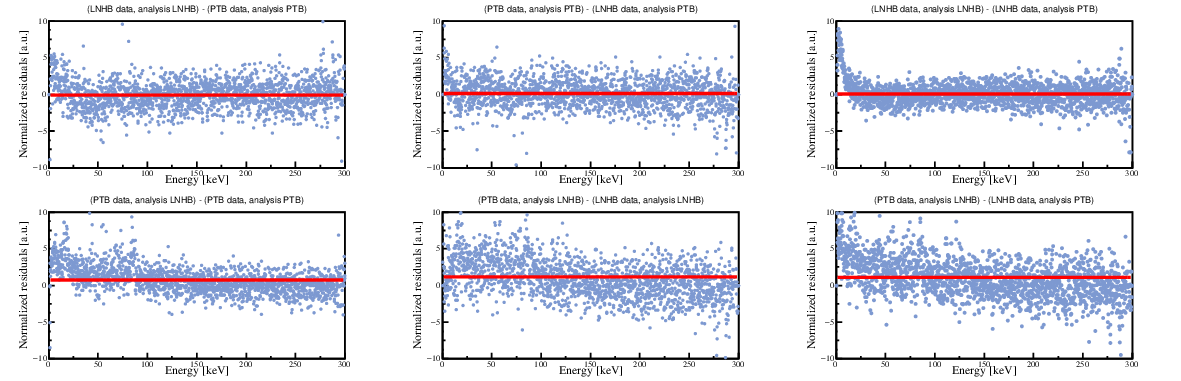}
        \caption{Comparison of the different analyses of the two MMC measurements of the $^{99}$Tc spectrum: normalized residual plots are shown for sets 3 and 2 (\textbf{top left}), sets 2 and 4 (\textbf{top center}), sets 3 and 4 (\textbf{top right}), sets 1 and 2 (\textbf{bottom left}), sets 1 and 3 (\textbf{bottom center}), sets 1 and 4 (\textbf{bottom right}).}
        \label{Fig:MMCComparison}
    \end{center}
\end{figure*}

After applying the corrections to the spectra as described, all four spectra should in principle show the same spectrum shape. Before these can be directly compared, they need to be normalized. We used the total number of events in the energy range from \SI{150}{\kilo\electronvolt} to \SI{250}{\kilo\electronvolt} as normalization factor, because there are no $\gamma$-lines in that range in either measurement. Normalized
residual plots of the various MMC sets, see Table \ref{tab:Cross_Analysis_Counts_Energy_Resolutions}, are depicted in Figure \ref{Fig:MMCComparison}.

\subsection{Silicon detectors}

The MMC results were compared to an independent measurement performed with a detection system recently developed at LNHB \cite{bisch2014etude, SinghThesis}. The system was designed for the measurement of beta spectra and is based on two PIPS detectors with thin entrance windows. Such detectors are commonly used to detect charged particles in Nuclear and Particle Physics.

\subsubsection{Experimental setup}

The geometry of the overall system is designed such that the PIPS detectors face each other. An ultra-thin radioactive source developed specifically for this application \cite{singh2019beta, singh2020experimental} is placed in their center. The experimental configuration covers more than 98\% of the solid angle and minimizes the self-absorption within the source. Consequently, experimental distortions of the spectrum are significantly reduced. Source and detectors are placed in an ultra-high vacuum chamber while the detection system is cooled down to \SI{100}{\kelvin} with liquid nitrogen, in order to improve the energy resolution and lower the detection threshold. The detector output is preamplified in the chamber before being shaped and recorded using a labZY module\cite{labZY2022}. A complete description of the detector design, acquisition system, and source fabrication can be found in \cite{Singh2023development}.

\subsubsection{Analysis and corrections}

The analysis of the data was performed in C++ using the ROOT software \cite{ROOTBRUN1997}. The system was calibrated with \rn{109}{Cd} and \rn{207}{Bi} sources whose emissions cover an energy range from about \SI{22}{\kilo\electronvolt} to \SI{1063}{\kilo\electronvolt}, as described in \cite{Singh2023development}. The energy resolution of the detection system was determined to be \SI{9}{\kilo\electronvolt} at \SI{65.52}{\kilo\electronvolt}, where an energy peak due to Ag K shell electrons which are emitted by internal conversion in \rn{109}{Cd} decay, is situated. The calibration sources were also used to determine the detection energy threshold which was estimated to be about \SI{10}{\kilo\electronvolt}.

The activity of the \rn{99}{Tc} source was about \SI{800}{\becquerel} in order to limit the dead time ratio of the acquisition to less than 0.5\%. The measurement was performed over five consecutive days. To estimate the contribution of the background to the spectrum, a ten-day measurement was performed with a source produced with a non-radioactive solution. The background was found to have little effect on the measured spectrum, as 80\% of the background events were detected below \SI{15}{\kilo\electronvolt}. For the data analysis, the background spectrum was subtracted from the \rn{99}{Tc} spectrum after normalization of the lifetime ratios of the measurements.

To account for the remaining distortions in the experimental spectrum, mainly due to energy loss in dead layers and escape of particles, an unfolding algorithm was adapted from the principle presented in \cite{Paulsen2020}. Based on detailed PENELOPE \cite{Salvat2015} Monte-Carlo simulations of the source-detector geometry, the algorithm provides a way to reconstruct the response matrix of the detection system. From this knowledge, the experimental spectrum can be unfolded to obtain the initial energy distribution of the beta electrons.  More details on the algorithm and the simulation can be found in \cite{Singh2023development}.
The response function of the detector is considered to be under good control from \SI{850}{\kilo\electronvolt} down to at least \SI{25}{\kilo\electronvolt}. This is evidenced by the good agreement between the analyzed data from several measured sources and the simulation of the detection system.

\subsubsection{Comparison}

\begin{figure}[ht]
    \begin{center}
        \includegraphics[width=\columnwidth]{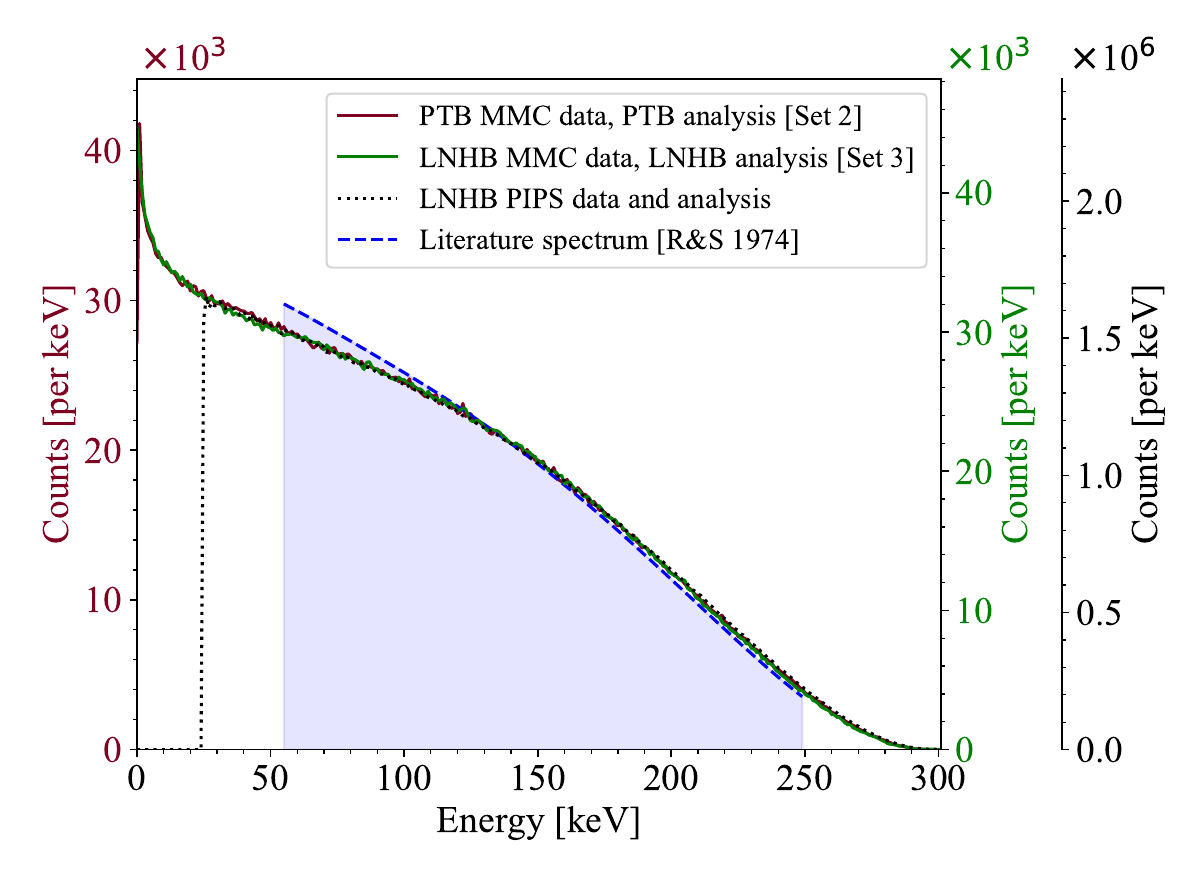}
        \caption{Comparison of the measured PTB MMC, LNHB MMC and PIPS spectra after removing the background in red, green and black, respectively. Please note the corresponding coloring of the first, second and third y-axis. Shown in blue is the recommended literature spectrum of \rn{99}{Tc} \cite{Reich74} which was plotted over the experimental measurement interval (\SI{55}{\kilo\electronvolt} to \SI{250}{\kilo\electronvolt}) using the reported shape factors. Notably, the literature spectrum shape diverges below \SI{100}{\kilo\electronvolt}.}
        \label{Fig:PTB_LNHB_MMCs_vs_LNHB_PIPS}
    \end{center}

\end{figure}

The corrected spectrum obtained from the PIPS measurement is compared in Fig. \ref{Fig:PTB_LNHB_MMCs_vs_LNHB_PIPS} to the two MMC measurements. The three spectra show an excellent agreement from the minimum reconstructed energy of the PIPS system up to the end point of the spectra. The PIPS measurement is independent of the MMC measurements, both in terms of detection method and source preparation, and the consistency between the three spectra underlines the reliability of the beta spectra presented in this article. Compared to the recommended beta spectrum of \rn{99}{Tc} in the literature \cite{Reich74}, which was reported by Reich and Schüpferling in 1974, there is rather good agreement above \SI{100}{\kilo\electronvolt} but divergence with a clear trend at lower energies.

\section{Combined analysis}
The results of Section \ref{subsec:E_max}, concerning the maximum beta energy were carried out at the PTB with theory insight from the LNHB and the spectrum-shape method calculations of Section \ref{sec:theoretical_study} were done at the LNHB.
\label{sec:combined_analysis}
\subsection{\texorpdfstring{\rn{99}{Tc}}{99Tc} \texorpdfstring{$Q$}--value}
\label{subsec:E_max}
For the determination of the maximum beta energy $E_{\mathrm{max}}$ of \rn{99}{Tc}, the methodology described in a previous article [\onlinecite{Kossert2022}] was adopted to allow for an energy-dependent shape-factor function $C(W)$. When neglecting the anti-neutrino mass, the beta spectrum is described by

\begin{eqnarray}\label{beta_spectrum}
    N(W)dW&=&\frac{G_{\beta}^2}{2\pi^3}F(Z,W)pW(W_0-W)^2dW\nonumber\\
          & & \cdot X(W)C(W)r(Z,W) ,
\end{eqnarray}

\noindent where $N(W)$ corresponds to the measured data; $F(Z,W)$ is the Fermi function with $Z$ the daughter atomic number; $p = (W^2-1)^{1/2}$; $W_0 = 1+E_{\mathrm{max}}/m_e$, where $m_e$ is the electron rest mass; $W = 1+E/m_e$; the constant $G_{\beta}^2$ is the squared product of the weak interaction coupling constant $g$ and the cosine of the Cabibbo angle, $\cos{\Theta_{\mathrm C}}$; and $X(W)$ stands for the correction of the atomic screening and exchange effects. The atomic overlap correction is given by

\begin{equation}
r(Z,W)=1-\frac{1}{W_0-W}\frac{\partial ^2}{\partial Z^2}B(G) .
\end{equation}

\noindent For \rn{99}{Tc} the constant $B^{''}=\frac{\partial ^2}{\partial Z^2}B(G)$ and its uncertainty was calculated to be $B^{''}$ = \SI{0.211(11)}{\kilo\electronvolt} when using parameterizations from the literature [\onlinecite{Hayen2018, Hardy2009}].

The shape-factor function of the $2^{\mathrm{nd}}$ forbidden non-unique beta transition of \rn{99}{Tc} is often parameterized as $1^{\mathrm{st}}$ forbidden unique using

\begin{equation}
C(W)=q^2+\lambda p^2=(W_0-W)^2+\lambda(W^2-1)
\end{equation}

\noindent with $\lambda$ being a constant parameter (see, e.g., \onlinecite{Reich74}). Hence, Eq. \ref{beta_spectrum} can be rearranged to get

\begin{eqnarray}\label{fit_function}
 \sqrt{\frac{N(W)}{pWF(Z,W)X(W)}}&=&K\bigg( \Big[ (W_0-W)^2  \nonumber \\
   & & - B^{''}(W_0-W)\Big]  \\
   & &\Big\{(W_0-W)^2+\lambda(W^2-1)\Big\}\bigg)^{1/2} ,\nonumber
\end{eqnarray}

\noindent which is then used for the fit procedures. In contrast to standard Kurie fits [\onlinecite{Kurie1936}] the fit function (right side of Eq. \ref{fit_function}) is not linear, but the three parameters ($K$, $W_0$ and $\lambda$) can be directly determined in a single fit process.

The analysis was carried out with all four MMC data sets (two measurements $\times$ two analyses). To this end, spectra with the background removed and corrected for bremsstrahlung losses and with a bin width of \SI{160}{\electronvolt} were used as starting point. If the used background model left the photon peaks of the external sources (\rn{57}{Co} and \rn{133}{Ba}, respectively) in place, these were removed. The uncertainty assigned to the background was conservatively estimated. To this end, the analysis was repeated without any background subtraction. The difference to the previous result with background subtraction was then used to evaluate the corresponding uncertainty component assuming a rectangular distribution. It should be noted that the influence of background to the determined maximum beta energy also depends on the energy range that is considered for the fits.\newline
\indent For all cases [sets 1-4], the spectrum unfolding shifted the maximum energy by approximately +\SI{200}{\electronvolt}. The impact on the spectrum shape, however, is very subtle as the energy losses of the absorbers were very small $<$1\% in these experiments. If the energy losses are larger, the unfolding effects the spectrum shape more significantly \cite{PaulsenPhD2022, Paulsen2020, SinghThesis}.

\begin{table}[!ht]
    \centering
    \caption{Results for the maximum beta energy $E_{\mathrm{max}}$ obtained using spectra from the two analysis codes and two measurements. The results correspond to mean values that were obtained from three individual results using different energy ranges for the fits (\SI{70}{\kilo\electronvolt} to \SI{290}{\kilo\electronvolt}, \SI{140}{\kilo\electronvolt} to \SI{290}{\kilo\electronvolt} and \SI{125}{\kilo\electronvolt} to \SI{293.5}{\kilo\electronvolt}, respectively). In all cases, the measurement uncertainties were taken into account (weighted fits), the background was subtracted, spectrum unfolding was taken into account and the shape-factor function $C(W)=q^2+\lambda p^2$ was assumed.}
    \begin{NiceTabular}{|c|c|c|c|c|}[hvlines,corners=NW,colortbl-like]
        \bf{Set}  & \bf{Measurement} & \bf{Analysis} & \bf{$E_{\mathrm{max}}$ in keV} & $\mathbf{\lambda}$\\
        1 & \Block[fill=burgundy!30]{1-1}{PTB} & \Block[fill=grannysmithapple!20]{1-1}{LNHB} & 295.809 & 0.615  \\
        2 & \Block[fill=burgundy!30]{1-1}{PTB} & \Block[fill=burgundy!30]{1-1}{PTB} & 295.786 & 0.652  \\
        3 & \Block[fill=grannysmithapple!20]{1-1}{LNHB} & \Block[fill=grannysmithapple!20]{1-1}{LNHB} & 295.845 & 0.651 \\
        4 & \Block[fill=grannysmithapple!20]{1-1}{LNHB} & \Block[fill=burgundy!30]{1-1}{PTB} & 295.862 & 0.659  \\
    \end{NiceTabular}
    \label{tab_emax_results}
\end{table}

\begin{table*}[ht]
  \centering
  \caption{Uncertainty budgets for the maximum beta energy $E_{\mathrm{max}}$ for the two measurements. All uncertainties are stated as standard uncertainties ($k = 1$).}
    \begin{NiceTabular}{|p{9.985em}|c|c|c|c|c|}
    \toprule
     & \Block[fill=grannysmithapple!20]{1-2}{\textbf{LNHB meas.}} & & \Block[fill=burgundy!30]{1-2}{\textbf{PTB meas.}}\\
    \textbf{Uncertainty} & \textbf{$u$} & \textbf{Relative } & \textbf{$u$} & \textbf{Relative } &  \\
    \textbf{component} & \textbf{in eV} & \textbf{uncertainty} & \textbf{in eV} & \textbf{uncertainty} & \textbf{Comment} \\\hline
    \textbf{Energy calibration} & 45    & 0.015\% & 50    & 0.017\% & \multicolumn{1}{p{24.715em}|}{Several well-known peaks used for the calibration, high reproducibility, different calibration sources in the two measurements (\rn{57}{Co} and \rn{133}{Ba})} \\ \hline
    \textbf{Resolution distortion effect, finite energy resolution} & 10    & 0.003\% & 10    & 0.003\% & \multicolumn{1}{p{24.715em}|}{Bins at high energies avoided} \\
    \hline
    \textbf{Background} & 141   & 0.048\% & 69    & 0.023\% & \multicolumn{1}{p{24.715em}|}{Background taken into account} \\
     \hline
    \textbf{Fit method} & 93    & 0.031\% & 89    & 0.030\% & \multicolumn{1}{p{24.715em}|}{Variation of energy range; weighted vs. unweighted fit, re-binning} \\
     \hline
    \textbf{Theoretical model} & 75    & 0.025\% & 75    & 0.025\% & \multicolumn{1}{p{24.715em}|}{Analysis with other shape-factor parameterization, difference with/without screening and exchange} \\
     \hline
     \textbf{Spectrum unfolding} & 57    & 0.019\% & 57    & 0.019\% & \multicolumn{1}{p{24.715em}|}{Dependence on cross sections (e.g. bremsstrahlung) and geometry} \\
     \hline

    \textbf{Analysis software/pile-up} & 10    & 0.003\% & 10    & 0.003\% & \multicolumn{1}{p{24.715em}|}{Deviation when using input spectra obtained from the two analysis codes} \\
     \hline
    \textbf{Combined} & 199   & 0.067\% & 156   & 0.053\% &  \\ \hline
    \end{NiceTabular}%
  \label{tab_emax_unc}%
\end{table*}

Results of the fit procedure are shown in Table \ref{tab_emax_results} and an uncertainty budget is shown in Table \ref{tab_emax_unc}. The evaluation of further uncertainty components was carried out in a similar manner as described in reference [\onlinecite{Kossert2022}]. For the uncertainty analysis fit ranges were varied, and unweighted fits were compared with fits that take statistical uncertainties into account. In order to evaluate a model uncertainty, the analyses were repeated ignoring the correction for screening and the atomic exchange effect and by using a modified shape-factor function
$C(W)=1+aW+b/W+cW^2$. In this case, the mean result agrees to within \SI{74}{\electronvolt}. However, a somewhat larger spread of results for the maximum energy is observed when using this shape-factor parameterization, which might be due to the larger number of adjustable parameters.
The uncertainty budgets for both measurements are listed in Table \ref{tab_emax_unc}.
The individual results $E_{\mathrm{max}} =$~\SI{295.798(156)}{\kilo\electronvolt} for the PTB measurement
and $E_{\mathrm{max}} =$~\SI{295.854(199)}{\kilo\electronvolt} for the LNHB measurement are used to calculate a weighted mean as our final result:
\begin{equation}
    E_{\mathrm{max}} = \SI{295.82(16)}{\kilo\electronvolt}.
\end{equation}
The uncertainty of the final result corresponds to the uncertainty of the PTB result and is more conservative than the inner (\SI{123}{\electronvolt}) and outer (\SI{3}{\electronvolt}) uncertainties of the weighted mean and we can exclude an underestimation of the uncertainty due to correlations.
The analysis described above also provides information on the parameter $\lambda$. However, the analysis is dedicated to the determination of the maximum energy, and it does not consider the low-energy part of the beta spectrum. Thus, the stated $\lambda$ values are not necessarily suited to describe the spectrum shape in a wide energy range.

	\subsection{The spectrum-shape method}
	\label{sec:theoretical_study}
In usual descriptions of nuclear beta decay, the Hamiltonian density is expressed in terms of lepton and hadron currents. Assuming a pure ($V-A$) weak interaction, it is convenient to introduce the ratio of the axial-vector coupling constant \ga to the vector coupling constant $g_V$. The Conserved Vector Current (CVC) hypothesis derives from the gauge invariance of the weak interaction and leads to $g_V = 1$. According to the Partially Conserved Axial-vector Current (PCAC) hypothesis, one can adopt the free-nucleon value \gafree = 1.2754 (13)~\cite{PDG2022}.\newline	
\indent An ideal description of the nuclear structure would allow the use of \gafree in beta decay calculations. However, the actual value of \ga can be renormalized in the decay because of the inevitable imperfections in any nuclear model. An effective value \gaeff then helps to compensate for some approximations like nonexistent or partial core excitations, or simplified many-nucleon correlations. This has been well known for a long time in the study of partial half-lives (see e.g. the review of Suhonen~\cite{Suhonen2017}). Recent theoretical studies suggested that \gaeff can also have a significant influence on the spectrum shape of forbidden non-unique transitions~\cite{Haa2016,Kos17_2}. Indeed, the calculation of forbidden non-unique transitions involves non-relativistic axial-vector matrix elements that are renormalized by the ratio $g_A$/$g_V$. In particular, \rn{99}{Tc} decay was predicted to be very sensitive to $g_A^{\text{eff}}$, making this spectrum a good candidate for a precise determination~\cite{Kostensalo2017}.\newline
\indent This approach, called the spectrum-shape method and introduced in~\cite{Haa2016}, was applied in the present work. The analysis was carried out on the spectrum measured and analyzed at PTB, chosen for its lower background correction. Because full computation of the beta spectrum is time consuming, the energy binning was increased to \SI{1}{\kilo\electronvolt}. The maximum energy used as input parameter was from this work, as described above.\newline
	\indent A formula is given in~\cite{Suhonen2017} to predict an effective value of \ga according to a quenching factor in infinite nuclear matter. Applied to $^{99}$Tc decay, we obtained \gainf = 1.120. As detailed below, the \gaeff value we extracted is far from this prediction.
	
\subsubsection{Beta spectrum modeling}
	
\begin{figure}[ht]
		\begin{center}
        \includegraphics[scale=0.44]{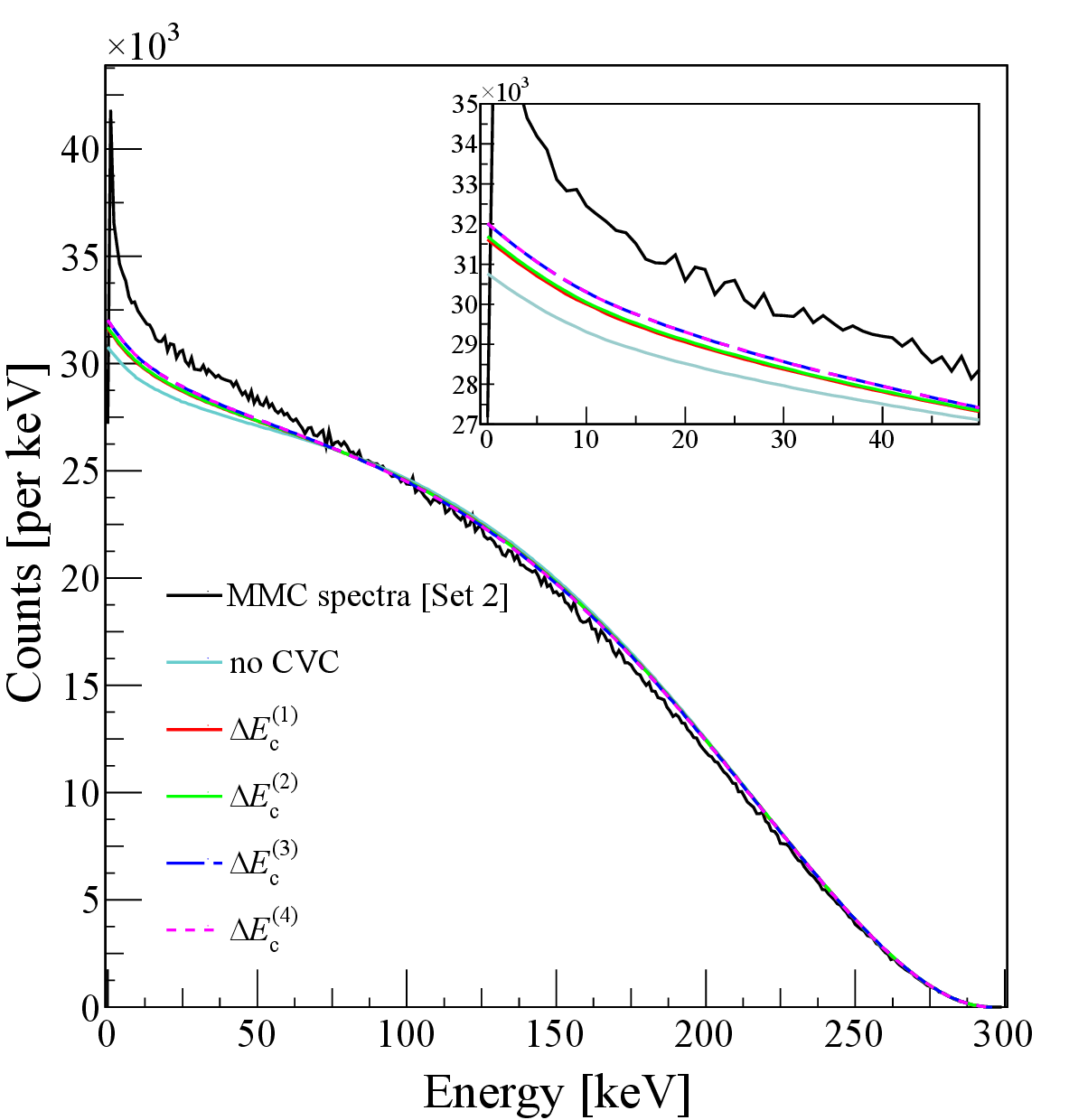}
			\caption{Comparison of the measured \rn{99}{Tc} spectrum with different theoretical curves. CVC is either ignored or included for different estimates of the Coulomb displacement energy $\Delta E_C$. An inset shows the low-energy part of the spectrum.}
			\label{fig:spectrum_CVC}
		\end{center}
	\end{figure}

The method to calculate the beta spectrum has already been described in~\cite{Kossert2022,Quarati2022} and follows the formalism of Behrens and B{\"u}hring~\cite{Behrens1982}. The theoretical spectrum is described by Eq. (\ref{beta_spectrum}) on which radiative corrections as detailed in [\onlinecite{Hayen2018}] are also applied.\newline	
\indent The shape factor $C(W)$ is a convolution of the nuclear structure and the lepton dynamics and is usually expanded in different multipoles of both the nuclear and lepton currents. The latter was simplified by Behrens and B{\"u}hring in order to decouple the calculation of lepton and nuclear matrix elements. The procedure consists in expanding the lepton radial wave function in powers\footnote{$R$ is the daughter nucleus radius and $\alpha$ is the fine structure constant.} of ($m_e R$), ($W R$) and ($\alpha Z$) and is here referred to as truncated lepton current. In the case of the second forbidden non-unique \rn{99}{Tc} decay, we followed Behrens' and B{\"u}hring's recommendation keeping only the dominant terms [\onlinecite{Behrens1982}]. It is noteworthy that in [\onlinecite{Suhonen2017}], the authors also considered next-to-leading order terms as described in~\cite{Haaranen2017}. This procedure is possible only with a simplified Coulomb potential. As in our previous study of \rn{151}{Sm} decay~\cite{Kossert2022}, we also considered a full numerical lepton current based on lepton wave functions determined by solving the Dirac equation with a Coulomb potential that includes atomic screening. All the terms of the lepton wave function expansion are therefore virtually accounted for.\newline	
\indent For calculating the shape factor, an input from a realistic nuclear structure model is required. The NuShellX code [\onlinecite{Brown2014}] was used in this work to determine the list of nucleon-nucleon transitions that contribute to the \rn{99}{Tc} decay. These single-particle transitions are weighted by their corresponding one-body transition densities. Following [\onlinecite{Kostensalo2017}], we first considered the effective interaction from Gloeckner [\onlinecite{Gloeckner1975}] with the \textit{GL} valence space spanning the proton orbitals $2p_{1/2}$ and $1g_{9/2}$, and the neutron orbitals $3s_{1/2}$ and $2d_{5/2}$. With such a description, the \rn{99}{Tc} decay is driven by a single nucleon-nucleon transition, from a neutron in $2d_{5/2}$ to a proton in $1g_{9/2}$. Next, the effective interaction from Mach [\onlinecite{Mach1990}] with the wider \textit{GLEKPN} valence space was considered. This valence space spans the proton orbitals $1f_{7/2}$, $1f_{5/2}$, $2p_{3/2}$, $2p_{1/2}$ and $1g_{9/2}$, and the neutron orbitals $1g_{9/2}$, $1g_{7/2}$, $2d_{5/2}$, $2d_{3/2}$ and $3s_{1/2}$. To limit the computational burden, the $1f_{7/2}$ proton and the $1g_{9/2}$ neutron orbitals were constrained to be full, and 4 protons were blocked in the $1f_{5/2}$ orbital. As a result, the \rn{99}{Tc} decay is still dominantly driven by the same single-particle transition but a small admixture of a transition from a neutron in $1g_{7/2}$ to a proton in $1g_{9/2}$ appears.\newline
\indent The transition probabilities of forbidden non-unique decay depend firstly on a relativistic vector matrix element. While a non-relativistic matrix element couples the large components of the nucleon wave functions, a relativistic matrix elements couples their small and large components together. However, most of the nuclear models, as NuShellX, are non-relativistic. An accurate estimate of this relativistic vector matrix element can be obtained from a non-relativistic vector matrix element employing the CVC hypothesis (see e.g. \rn{36}{Cl} decay in~\cite{Sadler1993}). In \rn{99}{Tc} decay, the relationship is
	\begin{equation}
		\tensor[^{\mathrm{V}}]{F}{_{221}} \simeq -\dfrac{R}{\sqrt{10}} \left[W_0 - (m_n - m_p) + \Delta E_C\right] \tensor[^{\mathrm{V}}]{F}{_{220}}
	\end{equation}
	with $m_n$ and $m_p$ the neutron and proton rest masses and \dEcb the Coulomb displacement energy. The small $Q$-value of \rn{99}{Tc} decay makes a critical good estimate of \dEcb because $\left[W_0 - (m_n - m_p)\right]$ = \SI{-0.489}{\mega\electronvolt} only.\newline
\indent As detailed in~\cite{Quarati2022}, the Coulomb displacement energy can be estimated from different methods. Usually, a uniformly charged sphere is considered and the expression only depends on the daughter nucleus through its atomic number and its nuclear radius, giving in the present case \dEc{1} = \SI{13.476}{\mega\electronvolt}. A close expression can be established that depends on both the parent and daughter atomic numbers and their nuclear radii, giving the second value \dEc{2} = \SI{12.814}{\mega\electronvolt}. However, \dEcb is known to possibly be sensitive to the mismatch between the initial and final nucleon wave functions~\cite{Damgaard1966}. Behrens and B{\"u}hring approximated the single-particle potential difference by the average of the Coulomb potential, keeping only the leading order of the lepton radial wave functions [\onlinecite{Behrens1982}]. This Coulomb displacement energy \dEc{3} is thus different for each nucleon-nucleon transition but is still independent of the beta-particle energy. For the dominant single-particle transition, we obtained the value \dEc{3} = \SI{10.560}{\mega\electronvolt}. Finally, we also introduced a dependency on the beta-particle energy by considering the complete lepton wave functions with numerical integration. The corresponding Coulomb displacement energy is denoted \dEc{4}, this method being in principle the most accurate. On average, \dEc{4} was found to be close to \dEc{3} for identical nucleon-nucleon transition.\newline
 \indent We present in Fig.~\ref{fig:spectrum_CVC} the spectrum calculated without the CVC hypothesis, and the spectra considering CVC for the different \dEcb estimates. The free-neutron value of \ga was assumed. CVC has an influence but does not lead to a change of the spectrum shape as spectacular as in \rn{36}{Cl} decay~\cite{Sadler1993}. The spectrum mostly exhibits a dependency at low energy and the spectra with \dEc{3} and \dEc{4} are hardly distinguished. In addition, we tried to adjust \dEcb in order to be as close as possible to the measured spectrum. The best value was found to be \dEcb = \SI{3.57}{\mega\electronvolt} with a poor reduced-$\chi^2$ of 5.018, and a non-linear tendency was found in the residuals. Most of all, it is clear that the adjusted Coulomb displacement energy is not realistic.\newline
\indent This result supports an adjustment of \ga to retrieve the measured spectrum. We chose what should be the most accurate theoretical spectrum as reference for the fitting procedure, i.e. considering a full numerical treatment of the lepton current, the one-body transition densities from the \textit{GLEKPN} valence space with Mach interaction, and the CVC hypothesis with the \dEc{4} Coulomb displacement energy.	
	\subsubsection{Effective \texorpdfstring{$g_A$}{gA} coupling constant}
\begin{figure}[ht]
		\begin{center}
            \includegraphics[scale=0.43]{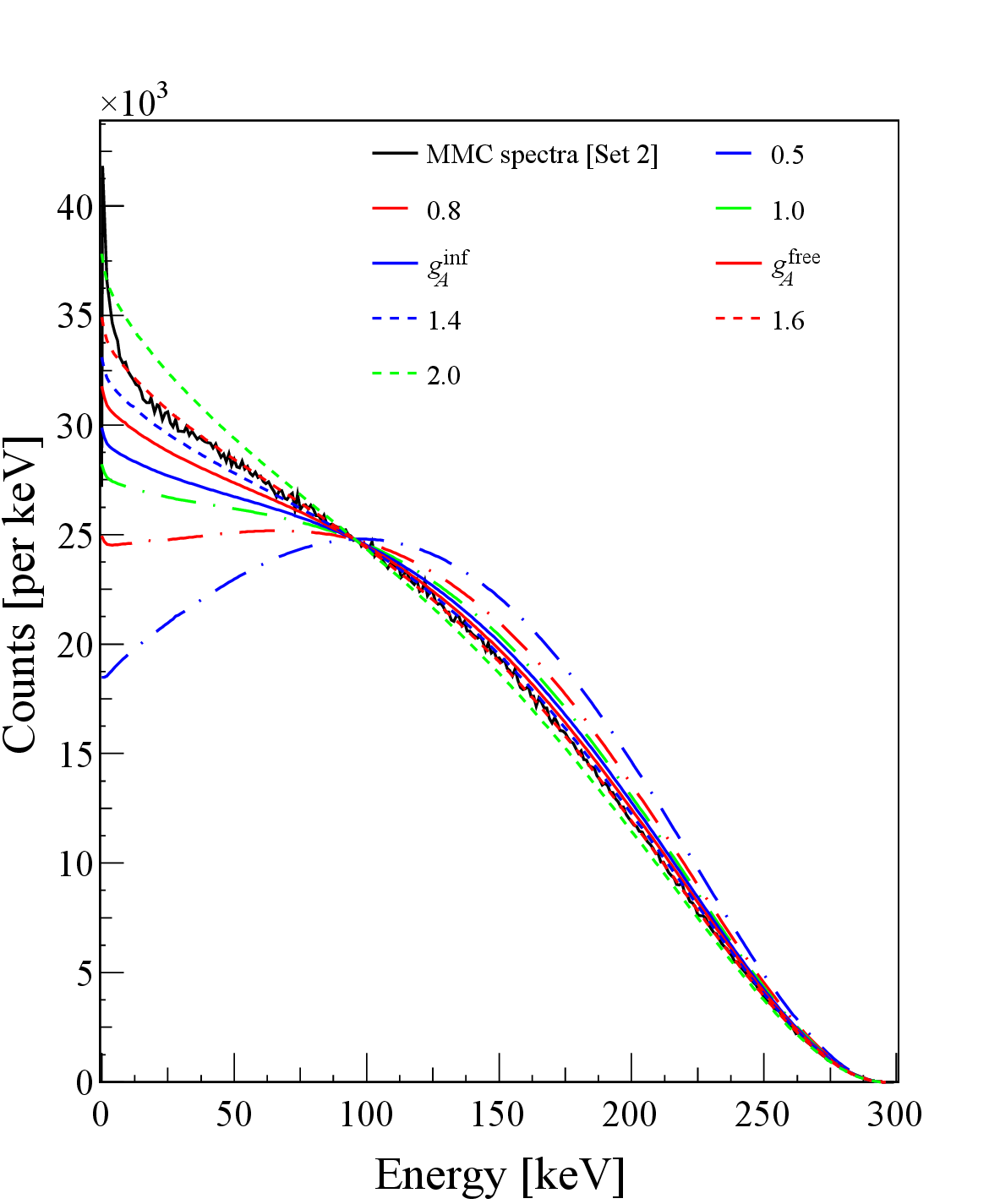}
			\caption{Influence of the effective \ga value on the theoretical beta spectrum of \rn{99}{Tc} decay, compared to the high-precision measurement from this work.}
			\label{fig:spectrum_gAeff}
		\end{center}
	\end{figure}

\begin{figure}[htbp]
		\begin{center}
   \includegraphics[scale=0.44]{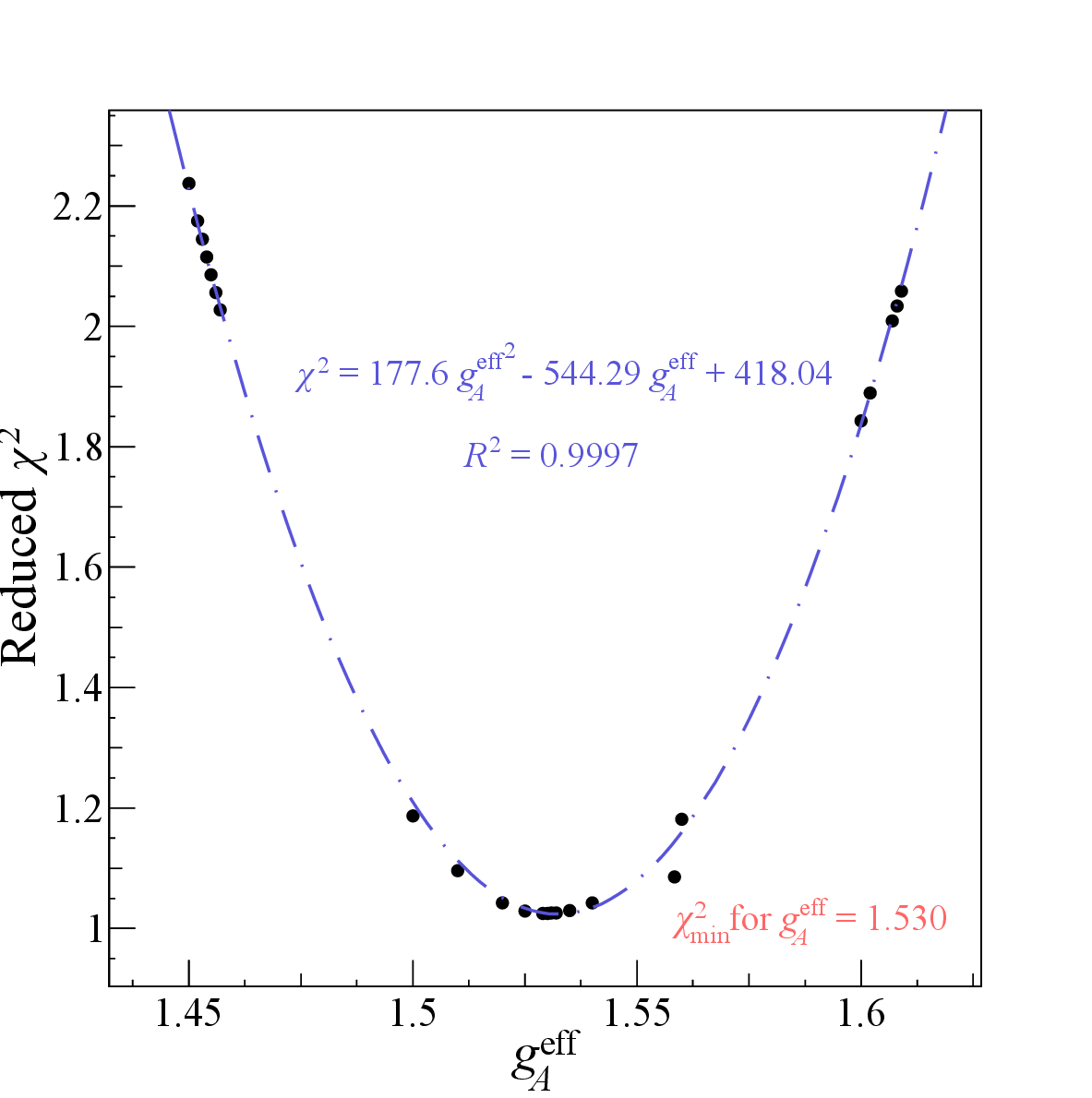}
			\caption{Reduced-$\chi^2$ distribution from the fitting procedure to extract \gaeff from the measured \rn{99}{Tc} spectrum. A quadratic curve has been fitted.}
			\label{fig:chi2_gAeff}
		\end{center}
	\end{figure}

Based on the modeling described above, the \ga value was varied from 0.5 to 2.0 to explore its influence. As illustrated in Fig.~\ref{fig:spectrum_gAeff}, we indeed observed a high sensitivity of the spectrum shape on the effective \ga value. However, we did not find the same behavior as in [\onlinecite{Kostensalo2017}] where shapes are similar at their extreme \gaeff values of 0.8 and 1.2, and are strongly different at \gaeff values of 1.0 and 1.1. If a simple typo in the curve labels in [\onlinecite{Kostensalo2017}] cannot be excluded, a possible explanation could be that these authors do not seem to consider the CVC hypothesis in their calculations. The sensitivity of our spectrum to \ga looks more consistent, with a probability at low energy that always increases for increasing \gaeff values.\newline
\indent Quick inspection of Fig.~\ref{fig:spectrum_gAeff} clearly shows that a \gaeff value between 1.4 and 1.6 could give good agreement with the measured spectrum from this work. The fit procedure simply consisted in redoing the calculations for different \gaeff values until minimum reduced-$\chi^2$ was found. A reasonable energy range of \SI{20}{\kilo\electronvolt} to \SI{275}{\kilo\electronvolt} was considered to determine the central value in order not to be influenced by the first and the last \SI{20}{\kilo\electronvolt} of the spectrum where distortions can be significant. The reduced-$\chi^2$ distribution is presented in Fig.~\ref{fig:chi2_gAeff} and was found to be very close to a quadratic shape. The best adjusted value is \gaeff = 1.530 with reduced-$\chi^2$ = 1.024.
	
\begin{figure}[ht]
		\begin{center}
            \includegraphics[scale=0.49]{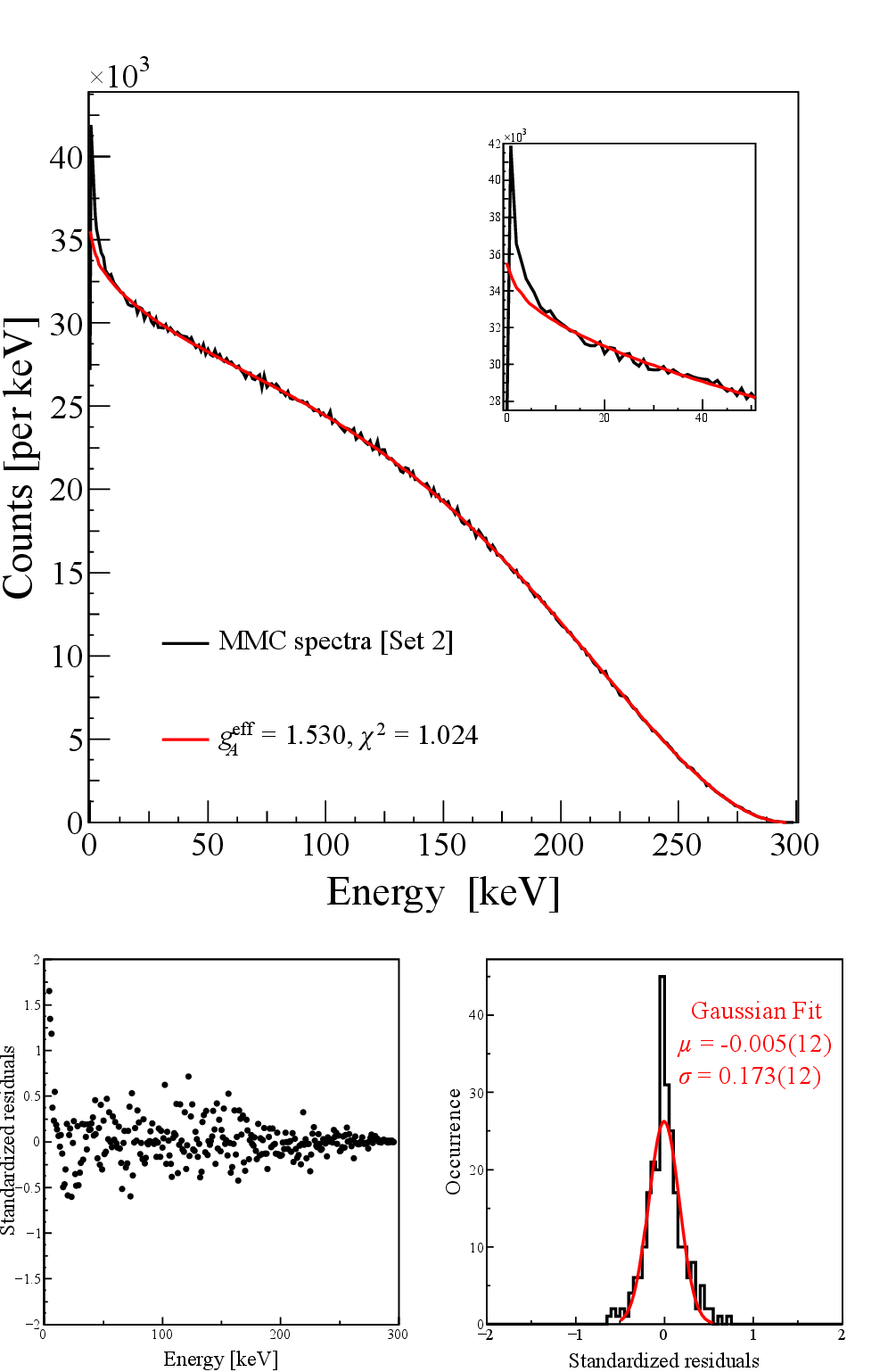}
			\caption{Comparison of the measured \rn{99}{Tc} spectrum with the theoretical curve with the best adjusted \gaeff value. An inset shows the low-energy part of the spectrum and the distributions of the residuals are also given.}
			\label{fig:spectrum_gAeff_fitted}
		\end{center}
	\end{figure}

	The final spectrum is compared to the high-precision measurement from this work in Fig.~\ref{fig:spectrum_gAeff_fitted}. The agreement is excellent down to \SI{6}{\kilo\electronvolt}. The distribution of the residuals does not show any energy dependency and follows a narrow Gaussian distribution, perfectly centered on zero. It is noteworthy that such agreement would not have been possible with an erroneous endpoint energy, which validates the extracted $Q$-value determined previously. Below \SI{6}{\kilo\electronvolt}, the atomic exchange correction does not seem to be sufficiently high to account for the distortion observed in the measured spectrum. We found the same discrepancy in the recent study of $^{151}$Sm decay [\onlinecite{Kossert2022}]. A possible explanation could be an inaccurate modeling of this atomic effect, as recently pointed out in [\onlinecite{Nitescu2023}]. These authors provide an analytical fit of their correction but only for allowed transitions. We applied it only to the Fermi function, which should give the main contribution. The very low-energy part of the spectrum is quite comparable and does not resolve the discrepancy with the measured spectrum. The best adjusted value was found to be \gaeff = 1.525 with reduced-$\chi^2$ = 1.041, which was used to estimate an uncertainty component as explained below.
	
	\begin{table*}[htbp]
		\centering
		\caption{Uncertainty budget for the \gaeff value extracted from the \rn{99}{Tc} beta spectrum measured in this work. All uncertainties are stated as standard uncertainties ($k = 1$).}
		\begin{tabular}{|p{16.em}|c|c|c|}
			\hline
			\multirow{2}{*}{\textbf{Uncertainty component}} & \multirow{2}{*}{\textbf{Value}} & \textbf{Relative} & \multirow{2}{*}{\textbf{Comment}} \\
			&  & \textbf{uncertainty} & \\\hline
			\textbf{Fit method} & 0.0755    & 4.97\% & \multicolumn{1}{p{24.715em}|}{Estimate at $\chi^2 \pm 1$; includes bin statistics component.} \\
			\hline
			\textbf{Fit range} & 0.0078    & 0.51\% & \multicolumn{1}{p{24.715em}|}{Largest deviation observed with extreme energy ranges.} \\
			\hline
			\textbf{Maximum energy} & 0.0043    & 0.28\% & \multicolumn{1}{p{24.715em}|}{Q-value and uncertainty from this work.} \\ \hline
			\textbf{Nuclear model} & 0.0135    & 0.88\% & \multicolumn{1}{p{26.em}|}{Deviation with GL model space and Gloeckner interaction.} \\
			\hline
			\textbf{Lepton current} & 0.0090   & 0.59\% & \multicolumn{1}{p{24.715em}|}{Deviation with simplified lepton current and \dEc{3}.} \\
			\hline
			\textbf{Coulomb displacement energy} & 0.0282    & 1.84\% & \multicolumn{1}{p{24.715em}|}{Largest deviation observed with \dEc{1}.} \\
			\hline
			\textbf{Atomic exchange} & 0.0020 & 0.13\% & \multicolumn{1}{p{24.715em}|}{Deviation with correction from [\onlinecite{Nitescu2023}].} \\
			\hline
			\textbf{Radiative corrections} & 0.0037    & 0.24\% & \multicolumn{1}{p{24.715em}|}{Conservative estimate, with or without including them.} \\
			\hline
			\hline
			\textbf{Combined} & 0.0828   & 5.42\% & \multicolumn{1}{p{24.715em}|}{Quadratic sum.} \\ \hline
		\end{tabular}%
		\label{tab:gaeff_unc}%
	\end{table*}
	
The uncertainty budget is detailed in Table~\ref{tab:gaeff_unc}. Several components were studied and estimated with the minimum-maximum method. Different probability distributions were considered depending on the uncertainty component, with the objective of being realistic and conservative.
	
The main contribution comes from the fit method, which also includes the spectrum statistics. The quadratic shape of the reduced-$\chi^2$ distribution allows to estimate the fit method uncertainty from the \gaeff values at reduced-$\chi^2$+1. The chosen energy range for the fitting procedure has also an influence. Two extreme cases were considered, namely \SI{0}{\kilo\electronvolt} to \SI{296}{\kilo\electronvolt} and \SI{100}{\kilo\electronvolt} to \SI{200}{\kilo\electronvolt}. The largest deviation was used with a triangular probability distribution.
	
The maximum energy of the spectrum determined in this work was used for the calculation and its uncertainty was propagated with calculations at $\pm1\sigma$, and considering a rectangular probability distribution. The influence of the atomic overlap correction on the extracted \gaeff value is insignificant.
	
For the other components, spectrum calculations were performed with different hypotheses, \gaeff was extracted and the largest deviation from \gaeff = 1.530 was considered with a triangular probability distribution. Uncertainty due to nuclear model was estimated considering GL model space and Gloeckner interaction [\onlinecite{Gloeckner1975}]. Usual simplified lepton current was employed to assess an uncertainty due to lepton current treatment. The different methods for determining the Coulomb displacement energy were tested and the maximum deviation was found with \dEc{1}. The \gaeff value determined with the atomic exchange correction from [\onlinecite{Nitescu2023}] was considered to estimate an uncertainty due to this correction.
	
	Finally, we considered an uncertainty component due to the radiative corrections. The latter include the emission of real soft photons from the internal bremsstrahlung process. This part of the correction on the beta spectrum assumes that these photons are lost and thus not detected. However, there is no doubt that they are partially reabsorbed by the detection system, especially low-energy photons. A conservative estimate was obtained by simply ignoring the radiative corrections in the spectrum calculation. It is more than an extreme case of possible photon reabsorption because additional corrections that do not come from internal bremsstrahlung are also ignored.
	
	The total combined uncertainty is given in Table \ref{tab:gaeff_unc}. The main component comes from the statistics of the measurement. The value of the effective axial-vector coupling constant extracted from \rn{99}{Tc} spectrum is eventually:
	\begin{equation}
		g_A^{\text{eff}} = 1.530 (83).
	\end{equation}

Finally, we determined the average energy of the beta spectrum:
\begin{equation}
\overline{E}_{\beta} = \SI{98.45(20)}{\kilo\electronvolt}
\end{equation}
and the corresponding $\log f$ value:
\begin{equation}
\log f = -0.47660 (22).
\end{equation}
With the partial half-life from \cite{DDEP_v6}, we obtained:
\begin{equation}
\log ft = 12.3478 (23)
\end{equation}
for the ground-state-to-ground-state transition in \rn{99}{Tc} decay.

Notably, the experimental shape factor from \cite{Reich74} leads to $\overline{E}_{\beta}$ = \SI{95.91(5)}{\kilo\electronvolt}, $\log f = -0.8785 (11)$ and $\log ft = 11.9458 (25)$, using the $Q$-value from this work.

\section{Conclusion}
\label{sec:conclusion}
In the work described here it is once again demonstrated that MMC measurements are excellently suited for determining both the shape and the maximum energy of beta spectra. The measurements are characterized not only by the high linearity and high energy resolution but also by the fact that very low detection thresholds can be reached. The good agreement of two almost independent measurements also in the cross-analysis of the data increases confidence in the obtained beta spectra. Since the background is the major uncertainty contribution of the obtained maximum beta energy, as shown in Table \ref{tab_emax_unc}, we conclude that experimental design and optimized calibration sources are crucial to improve upon such measurements using MMCs.\newline
\indent
High quality PIPS detector measurements confirm the shape of the spectrum above \SI{25}{\kilo\electronvolt}.
We do not make a detailed comparison with previous determinations of the shape of the beta spectrum, but point out that parameterizations of the beta spectrum found in the literature must now be considered obsolete.
It should be noted that all previous measurements of the \rn{99}{Tc} beta spectrum had significantly higher energy thresholds, so that a significant part of the spectrum at low energies had to be considered as being unknown.
Combining our measurements with detailed theoretical calculations, we extracted new decay data of interest: $Q_{\beta}$ = \SI{295.82(16)}{\kilo\electronvolt},

$\overline{E}_{\beta} = 98.45 (20)~\text{keV}$, $\log f = -0.47660 (22)$ and $\log ft = 12.3478 (23)$. The spectrum shape was found to be very sensitive to the effective value of the axial-vector coupling constant, with $g_A^{\text{eff}} = 1.530 (83)$ giving the best agreement with our measurement.\newline
\indent
Our $Q$-value is five times more precise than the recommended one \cite{Wang2021} and shifted by +\SI{2}{\kilo\electronvolt}. The uncertainty is competitive with Penning trap measurements and we call for a confirmation of our $Q$-value using this method. Beyond MMCs, the active ACCESS (Array of Cryogenic Calorimeters to Evaluate Spectral Shapes) project aims to measure of forbidden beta decays such as \rn{99}{Tc} using an array of Neutron Transmutation-Doped germanium (Ge-NTD or NTD) detectors \cite{pagnanini2023array}. Our results could then be confirmed in the near future with another independent technique.

\bigskip
\begin{acknowledgments}
This work is part of the project 20FUN04 PrimA-LTD that has
received funding from the EMPIR programme co-financed by the
Participating States and from the European Union’s Horizon 2020
research and innovation programme. The Linux-Compute-Cluster at the PTB Berlin was used to run EGSnrc. We thank Gerd Lindner and Andreas Lübbert for their helpful cooperation.
\end{acknowledgments}

\bibliography{Tc99_MMC}

\end{document}